 \documentclass[pra,amsmath,amssymb,onecolumn, showpacs, superscriptaddress,10pt]{revtex4-1}

\usepackage{amsmath}
\usepackage{hyperref}
\usepackage{graphicx}
\usepackage{amsfonts}
\usepackage{amsthm}
\usepackage{cases}
\usepackage{bm}

\usepackage[normalem]{ulem}

\usepackage{color}
\definecolor{Blue}{rgb}{0.00, 0.00, 1.00}
\definecolor{Red}{rgb}{1.00, 0.00, 0.00}
\definecolor{Green}{rgb}{0.00, 1.00, 0.00}
\definecolor{Cyan}{rgb}{0.00, 0.70, 0.70}

\hypersetup{
    colorlinks=true,       
    linkcolor=red,          
    citecolor=blue,        
    filecolor=magenta,      
    urlcolor=cyan           
}

\newcommand{\nn}{\nonumber}
\newcommand{\be}{\begin{equation}}
\newcommand{\ee}{\end{equation}}
\newcommand{\bea}{\begin{eqnarray}}
\newcommand{\eea}{\end{eqnarray}}

\newcommand{\beq}{\begin{equation}}
\newcommand{\eeq}{\end{equation}}
\newcommand{\beqn}{\begin{eqnarray}}
\newcommand{\eeqn}{\end{eqnarray}}

\begin{document}

\title{Stationary nonequilibrium bound state of a pair of run and tumble particles}

\author{Pierre Le Doussal}
\affiliation{CNRS-Laboratoire de Physique Th\'eorique de l'Ecole Normale Sup\'erieure, 24 rue Lhomond, 75231 Paris Cedex, France}
\author{Satya N. \surname{Majumdar}}
\affiliation{Universit{\'e} Paris-Saclay, CNRS, LPTMS, 91405, Orsay, France}
\author{Gr\'egory \surname{Schehr}}
\affiliation{Sorbonne Universit\'e, Laboratoire de Physique Th\'eorique et Hautes Energies, CNRS UMR 7589, 4 Place Jussieu, 75252 Paris Cedex 05, France}
%

\date{\today}

\begin{abstract}
We study two interacting identical run and tumble particles (RTP's) in one dimension. Each particle is driven by a telegraphic noise, and in some cases, also subjected to a thermal white noise with a corresponding diffusion constant $D$. We are interested in the stationary bound state formed
by the two RTP's in the presence of a mutual attractive interaction. The distribution of the relative coordinate $y$ indeed reaches a steady state that we characterize in terms of
the solution of a second-order differential equation. We obtain the explicit formula for the stationary probability $P(y)$ of $y$ for two examples of interaction potential $V(y)$. The first one corresponds to $V(y) \sim |y|$. In this case, for $D=0$ we find that $P(y)$ contains a delta function part at $y=0$, signaling a strong clustering effect, 
together with a smooth exponential component. For $D>0$, the delta function part broadens, leading instead to weak clustering. The second example is the harmonic attraction $V(y) \sim y^2$ in which case, for $D=0$, $P(y)$ is supported on a finite interval. We unveil an interesting relation between this two-RTP model with harmonic attraction and a three-state single RTP model in one dimension, as well as with a four-state single RTP model in two dimensions. We also provide a general discussion of the stationary bound state, including examples where it is not unique, e.g., when the particles cannot cross due to an additional short-range repulsion. 
 \end{abstract}

\maketitle

\tableofcontents

%
%

\section{Introduction}

Interacting active particles is a subject of much current interest both theoretically and experimentally~\cite{soft,BechingerRev,Ramaswamy2017,Marchetti2018,Berg2004,Cates2012,TailleurCates}. An active particle, in contrast to a passive particle, has
an autonomous self-propelled motion, which is modelled by a driving ``active" noise, which has a finite persistence time. For example, a commonly 
studied model is the so-called run-and-tumble particle (RTP) -- a motion exhibited by E. Coli bacteria \cite{Berg2004,TailleurCates}. In this simplest RTP model, the particle chooses
a direction at random and moves ballistically with a constant speed $v_0$ in that direction during an exponentially distributed random run time with mean $\gamma^{-1}$. 
Then it tumbles, i.e., it changes its direction at random and again moves ballistically with speed $v_0$, performing a new run. Thus runs and tumbles alternate. The tumbling rate $\gamma$ and the speed $v_0$ are the two parameters in this simplest RTP model. For example, in one dimension, the position $x(t)$ of the RTP evolves via the stochastic equation
\bea \label{def_RTP}
\frac{dx(t)}{dt} = v_0 \, \sigma(t) \;,
\eea
where $\sigma(t)$ is a telegraphic noise that takes values $\sigma(t) = \pm 1$ and changes from one state to another with a constant rate $\gamma$. Thus this ``active noise" $\sigma(t)$ 
has zero mean $\langle \sigma(t) \rangle = 0$ and its auto-correlation function decays exponentially in time $\langle \sigma(t) \sigma(t') \rangle = e^{-2\gamma\, |t-t'|}$. Therefore the active noise is
non-Markovian since it has a finite memory characterised by the persistence time $\gamma^{-1}$. In fact, much before the current interest in the context of active matter, this RTP model in one-dimension has been studied extensively both in the mathematics and the physics literature where it is known as ``persistent'' random walk~\cite{kac74,Orshinger90,W02,HJ95,ML17}. In the limit $\gamma \to \infty$, the active noise reduces to a ``passive'' delta-correlated noise. 
At long times, the effect of activity becomes somewhat insignificant since a free RTP is known to converge to a Brownian motion with an effective diffusion constant $D_{\rm eff} = v_0^2/(2 \gamma)$.
Thus the presence of activity is detected only in the effective diffusion constant $D_{\rm eff}$. One can also add a thermal noise in Eq. (\ref{def_RTP}) 
\bea \label{def_RTP_D}
\frac{dx(t)}{dt} = v_0 \, \sigma(t) + \sqrt{2\,D} \xi(t) \;,
\eea
where $\xi(t)$ is a Gaussian white noise with zero mean and a correlator $\langle \xi(t) \xi(t')\rangle = \delta(t-t')$. Here also the system becomes diffusive at late times with an effective diffusion constant $D_{\rm eff} = v_0^2/(2 \gamma) + D$ \cite{BechingerRev,MJK18}. Thus the effect of an additional thermal noise, in this simple setting, is just to renormalize the effective diffusion constant at late times. 

There are two natural generalisations of this single free RTP dynamics described above in (\ref{def_RTP}). The first concerns the long-time stationary state of the RTP in the presence of an external confining potential. In this case, the evolution equation (\ref{def_RTP_D}) has an additional external force $F(x) =-U'(x)$, with $U(x)$ being the confining potential, 
\bea \label{RTP_potential}
\frac{dx(t)}{dt} = F(x)\, + \, v_0 \, \sigma(t) + \sqrt{2\,D} \xi(t) \;.
\eea
Here, at late times, the system reaches a stationary state which is typically non-Boltzmann, thus retaining the effect of activity even at late times \cite{HJ95,Solon15, TDV16, DKM19, BMR19, DD19,3statesBasu,LMS2020}. 
The second generalisation is to study several RTP's with pairwise interactions (repulsive or attractive) between them. In the presence of interactions, RTP's are known to exhibit interesting collective effects, such as clustering and jamming~\cite{soft,Ramaswamy2017,TailleurCates,slowman, slowman2}. While there have been several studies on the effect of interactions between RTP's, there still are very few exact results available. For example, even for two RTP's on a ring with hard-core repulsion between them, the steady state exhibits clustering and the solution is nontrivial \cite{slowman,slowman2,KunduGap2020,MBE2019}. Note that with repulsive interactions between the RTP's, the steady state will exist only in a finite size system. In an infinite system with two particles, while there is no steady state, other dynamical properties have been studied -- for example the probability that two particles do not cross each other up to time $t$ has been 
computed exactly and it was shown to be already nontrivial due to the presence of the active noise \cite{LMS2019}.

To obtain a steady state for a system of RTP's in an infinite system (subtracting the zero-mode if any), one needs to 
introduce an attractive interaction between the RTP's. For instance, in the simplest setting of two particles with attractive interactions between them, one would expect to see clustering 
in the steady state in the form of a bound pair. The stationary properties of such bound pairs, even in an infinite system, are difficult to describe analytically. In fact, there are hardly any analytical result available in the literature on such bound pairs. 

\begin{figure}[t]
    \centering
    \includegraphics[width = 0.8 \linewidth]{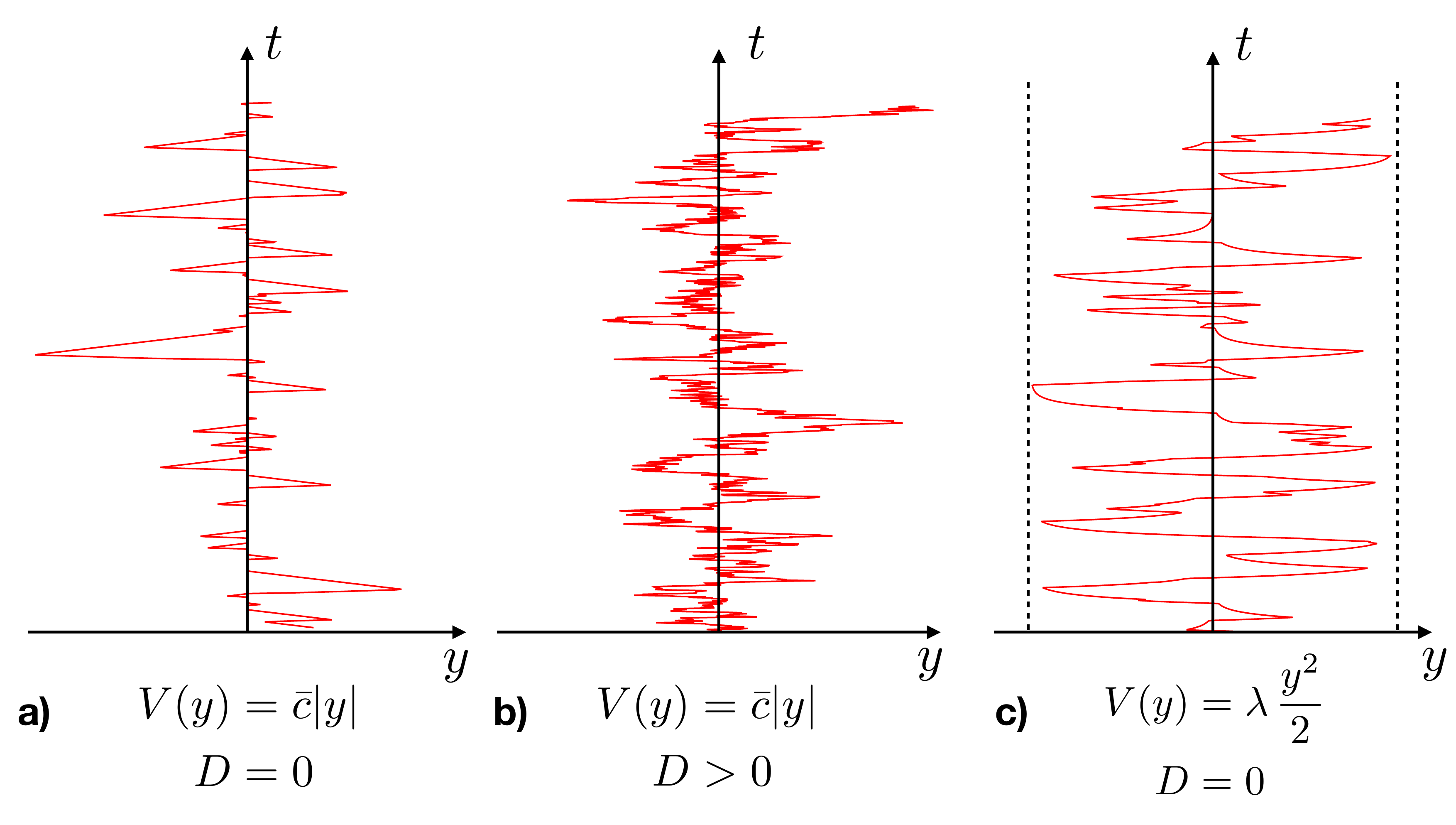}
    \caption{Typical trajectories of the relative coordinate $y$ as a function of time $t$ (obtained by solving numerically the equation of motion (\ref{relative0})) for the three different models for which we compute here exactly the stationary distribution $P(y, t \to \infty)$: {\bf a)} $V(y) = {\bar c}|y|$ with $D=0$, {\bf b)} $V(y) = {\bar c}|y|$ and $D>0$ and {\bf c)} $V(y) = \lambda y^2/2$ with $D=0$. In case {\bf a)}, the system exhibits strong clustering characterized by the presence of a delta-function at $y=0$ in the steady state [see Eq. (\ref{tot_prob})] and clearly seen on the figure where $y$ sticks to zero from time to time, while in case {\bf b)} the system exhibits only weak clustering since the thermal noise (i.e., $D>0$) broadens the delta-function [see Eq. (\ref{fullP_Dpos})]. Finally in case {\bf c)}, the relative coordinate $y$ is bounded in the steady state, as indicated by the vertical dotted lines [see e.g., Eq.~(\ref{exact_quad})].}
    \label{fig:traj}
\end{figure}

In this paper, we study a simple model of two RTP's on an infinite line with attractive interaction between them. We provide exact analytical results for the steady-state distribution 
of the inter-particle distance for different types of attractive interactions. These results provide a complete characterization of the bound pair. Even though our system is extremely simple, it turns
out that the stationary state of the bound pair has a very rich structure which depends on the shape of the interaction. In some cases, e.g., for a linear interaction potential, and 
when the particles are driven purely by active noise [see Fig. \ref{fig:traj} a)], one finds that the clustering is ``strong", a signature of which is the presence of a Dirac delta-function in the steady-state distribution of the inter-particle distance. In that case, adding the thermal noise broadens the delta-function, indicating a ``weak" clustering [see Fig. \ref{fig:traj} b)], with exponential decay of the steady-state distribution of the inter-particle distance. In other cases, e.g. for a harmonic interaction potential, the inter-particle distance in the steady state remains bounded in a finite interval [see Fig. \ref{fig:traj} c)]. 

The rest of the paper is organised as follows. In Section \ref{sec:model}, we introduce precisely our model and summarise the main results. In Section \ref{sec:lin}, we focus on the special case of a linear attractive potential $V(y) = \bar c |y|$ for which we compute exactly the stationary state, both in the absence (Section \ref{sec:linD0}) and in the presence (Section \ref{sec:linDnon0}) of the thermal noise. In Section \ref{sec:gen}, we study the case of a general $V(y)$, but in the absence of thermal noise. Finally we conclude in Section \ref{sec:conclusion}. 

\section{The model and a summary of the results} \label{sec:model}

In this paper, we consider two interacting RTP's on the line described by the equation of motion
\begin{equation}\label{langevin}
\frac{dx_1}{dt}= f(x_1-x_2)  + v_0\, \sigma_1(t) + \sqrt{2 D}\, \xi_1(t) \, , \quad \frac{dx_2}{dt}= f(x_2-x_1)  + v_0 \sigma_2(t) + \sqrt{2 D} \, \xi_2(t) \, \quad , \quad f(-y)=-f(y) \;,
\end{equation}
where $\sigma_1(t)=\pm 1$ and $\sigma_2(t)=\pm 1$ are two independent telegraphic noises
with flipping rate $\gamma$. In addition, both particles are also driven by thermal noises, represented by independent Gaussian white noises $\xi_1(t)$ and $\xi_2(t)$
of zero mean and correlators $\langle \xi_i(t) \xi_j(t') \rangle = \delta_{ij} \delta(t-t')$. We assume that both of them have the same diffusion constant $D$. 
The two RTP's interact via a potential energy $V(x_1-x_2)$ and in Eq. \eqref{langevin} $f(y) = - V'(y)$ is the inter-particle force. 
Equivalently, denoting $w=(x_1+x_2)/2$ and $y=x_1-x_2$ one has
\begin{eqnarray} 
\frac{dw}{dt}&=&  \frac{v_0}{2}\, (\sigma_1(t) + \sigma_2(t)) + \sqrt{D} \, \tilde \eta(t) \;, \label{cofm}  \\
\frac{dy}{dt}&=& 2 f(y) + v_0\, (\sigma_1(t) - \sigma_2(t) ) + \sqrt{4 D} \, \eta(t)  \label{relative0}  \;,
\end{eqnarray}
where $\eta(t) = (\xi_1(t)-\xi_2(t))/\sqrt{2}$ and $\tilde \eta(t)=(\xi_1(t)+\xi_2(t))/\sqrt{2}$ are two independent Gaussian white noises with zero mean, each with a delta-correlator.
The center of mass $w$ undergoes a free RTP motion similar to Eq. (\ref{def_RTP_D}) and clearly does not reach a stationary state. 
Hence we will focus here only on the relative coordinate $y(t)$ which evolves independently of $w(t)$. In other words
we study the system in the center of mass frame. By comparing Eqs. (\ref{RTP_potential}) and (\ref{relative0}), we see that the dynamics of the relative coordinate $y(t)$ in Eq. (\ref{relative0}) can 
be interpreted as the dynamics of the position of 
a single RTP with three internal states $(-2 v_0, 0, 2 v_0)$, and subjected to an external force $2 f(y)$. If $f(y)$ is sufficiently attractive we expect
that the relative coordinate will reach a stationary state, leading to a stationary bound
state of the pair of particles. Denoting by $P(y,t)$ the distribution of the relative coordinate at time $t$, our goal is to evaluate its stationary limit $P(y,t \to \infty)$.  
Note that the only difference between Eqs. (\ref{RTP_potential}) and (\ref{relative0}) is that the driving active noise in the former case has two states, while in the latter it
has three states. In the former case (\ref{RTP_potential}), the stationary position distribution for arbitrary $f(y)$ is exactly known, at least for $D=0$. In contrast, 
when the driving active noise has three states, as in Eq. (\ref{relative0}), it is more difficult to compute the stationary distribution of the relative coordinate for arbitrary $f(y)$, even 
for $D=0$. In this paper we derive a second order differential equation, see \eqref{final}, which is obeyed by this stationary distribution $P(y,t \to \infty)$. It is challenging to
solve it analytically for a general $f(y)$, but here we obtain explicit solutions for two special cases of $f(y)$ as discussed below.

To compute the stationary state, we start from the Langevin equation (\ref{relative0}) for the relative coordinate $y(t)$, and write down the corresponding Fokker-Planck (FP) equation. However, due to the 
non-Markovian nature of the active noises $\sigma_1(t)$ and $\sigma_2(t)$, the process is Markov only when one keeps track of the two internal degrees of freedom
$\sigma_1(t)$ and $\sigma_2(t)$, in addition to $y(t)$. This obliges us to define $P_{\sigma_1, \sigma_2}(y,t)$ as the probability density for the relative coordinate to be
at $y$ at time $t$ and that the internal ``spins" $\sigma_1(t)$ and $\sigma_2(t)$ take values $\sigma_1$ and $\sigma_2$ at time $t$. One can then write down the four
coupled FP equations for $P_{\sigma_1, \sigma_2}(y,t)$ corresponding to $\sigma_1 = \pm 1$ and $\sigma_2 = \pm 1$. The distribution of the relative coordinate $P(y,t)$ is then
obtained by summing over the four internal states
\bea \label{P_4states}
P(y,t) = \sum_{\sigma_1 = \pm 1, \sigma_2 = \pm 1} P_{\sigma_1, \sigma_2}(y,t) \;.
\eea
However, for general $f(y)$, solving these
FP equations, even in the stationary state, turns out to be rather hard. There is however one special case, namely when $V(y) = \bar{c}\, |y|$ (corresponding to $f(y) = - \bar{c}\,{\rm sgn}(y)$), 
for which one can obtain the stationary state explicitly for arbitrary $D\geq 0$. We present this solution in detail in Section \ref{sec:lin}, first in the absence of the thermal noise ($D=0$) and
then in the presence of the thermal noise $D>0$. This allows us investigate the effect of thermal noise on the stationary state. Our main result is that, when $D=0$, the stationary state $P(y,t \to \infty)$ has two parts: (i) an exponentially decaying part and (ii) a delta-function at $y=0$ [see Eqs. (\ref{result2}) and (\ref{res_simple})]. As mentioned above, the presence of this delta-function is a signature of {\it strong} clustering. When $D>0$ is switched on, the delta-function part gets smeared and the stationary state consists only of decaying exponentials [see for instance Eq.~(\ref{fullP_Dpos})]. Thus the effect of the thermal noise in the stationary state is to weaken the clustering. The second case for which we find an explicit solution is the harmonic potential $V(y) = \lambda y^2/2$,
which we study only for $D=0$. In that case, the support of the stationary distribution of the relative coordinate is a single interval, where $P(y)$ exhibits some power-law singular behavior near the 
edges and at the center $y=0$, with exponents depending on the parameters $\gamma$ and $\lambda$. It turns out that in this harmonic case the stationary solution is identical to the solution of another three-state model that was studied recently \cite{3statesBasu}, even though the dynamics of the two models are quite different. In the presence of a $D>0$, although we did not study it, we expect the support to extend to the full real axis.

\section{Stationary solution for the linear interaction potential} \label{sec:lin}

In this section we will study the stationary state of the inter-particle distance $y(t)$ in the presence of an attractive potential $V(y)= \bar c |y|$,
with $\bar c>0$ (similar to the Coulomb interaction between two opposite charges in one dimension). This corresponds to a force
$f(y) = - V'(y) = - \bar c \, {\rm sgn}(y)$. In this case, the evolution equation 
for $y(t)$ in Eq. (\ref{relative0}) reads
\bea \label{relative_D0}
\frac{dy}{dt}= - 2 \bar c \, {\rm sgn}(y)  + v_0\, (\sigma_1(t) - \sigma_2(t) ) + \sqrt{4 D} \, \eta(t)  \;.
\eea
In general, for arbitrary $f(y)$ in Eq. (\ref{relative0}), it is not easy to compute the stationary state in the presence of the thermal noise ($D>0$). However,
in this special case when $f(y) = - \bar c \, {\rm sgn}(y)$, we show below that the stationary state for $y(t)$ can be fully characterised, both  for $D=0$ and for $D > 0$.

\subsection{Without thermal noise, $D=0$} \label{sec:linD0}

In this section we study the process in Eq. (\ref{relative_D0}) in the absence of thermal noise $D=0$.
We note that $y(t)$ in Eq. (\ref{relative_D0}) is actually a non-Markov process, since $\sigma_1(t)$ and $\sigma_2(t)$
have a finite memory. In order to write a FP equation, we need to recast first the dynamics into a Markovian
form. This is usually done by enlarging the phase space -- here, e.g., by considering the evolution of the triplet $\{y(t), \sigma_1(t), \sigma_2(t) \}$. 
This leads us to define $P_{\sigma_1,\sigma_2}(y,t)$ as the probability density function (PDF) of the relative
coordinate at time $t$ with internal states $\sigma_1,\sigma_2$. The time evolution of this PDF
is governed by the following FP equation
\bea \label{FP}
\partial_t P_{\sigma_1,\sigma_2} = - \partial_{y} [ (- 2 \bar c \,  {\rm sgn}(y) + v_0 
(\sigma_1-\sigma_2) ) P_{\sigma_1,\sigma_2} ] - 2 \gamma P_{\sigma_1,\sigma_2} + \gamma (P_{-\sigma_1,\sigma_2} + P_{\sigma_1,-\sigma_2}) \;.
\eea 
In Eq. (\ref{FP}), $\sigma_1$ and $\sigma_2$ can both take values $\pm 1$. Hence, Eq. (\ref{FP}) describes actually four coupled equations depending on the four values of $\{\sigma_1, \sigma_2 \}$, namely $P_{++}(y,t), P_{+-}(y,t), P_{-+}(y,t)$ and $P_{--}(y,t)$. The first term in the right hand side (r.h.s.) of Eq. (\ref{FP}) describes the convection in the presence of an external force,
while the rest of the terms denote the loss and gain due to the flipping of the telegraphic noise.
The total probability $P(y,t)$ is then obtained by summing over the internal degrees of freedom as in Eq. (\ref{P_4states}). 

Before analyzing the FP equation \eqref{FP} let us investigate the Langevin equation and see what we
may anticipate for the evolution of the system. It reads 
\be \label{Langeviny}
\frac{dy}{dt}= - 2 \, \bar c \, {\rm sgn}(y)  + v_0\, (\sigma_1(t) - \sigma_2(t) ) \;.
\ee
Consider for instance the case when $(\sigma_1(t),\sigma_2(t))$ are either $(+,+)$ or $(-,-)$, in which case
$\frac{dy}{dt}= - 2 \bar c \, {\rm sgn}(y)$. Therefore, for $y(0)>0$ the time evolution is $y(t)=y(0)- 2 \bar c t $
and $y(t)$ vanishes in finite time. At all later times it remains zero until one of the $\sigma_i(t)$
changes sign provided $v_0 > \bar c$. This is a clustering effect which will lead to the appearance of a delta function component, $\propto \delta(y)$ 
in $P_{\sigma_1,\sigma_2}(y,t)$. In fact when $v_0 < \bar c$ we expect that the total probability $P(y,t)$
converges to $\delta(y)$ in finite time, and remains there. In contrast for $v_0 > \bar c$ we
expect a non trivial stationary distribution, where the delta function at $y=0$ coexists with a continuous
background. 

We expect the system to reach a stationary state in the long time limit $t \to \infty$. For simplicity of notations, we will denote the stationary state by $P_{\sigma_1,\sigma_2}(y) = P_{\sigma_1,\sigma_2}(y, t \to \infty)$. The stationary solution can be obtained from Eq. (\ref{FP}) by setting $\partial_t P_{\sigma_1,\sigma_2}=0$ in
the left hand side (l.h.s.) of Eq. (\ref{FP}). This leads to
\be
\label{FPS}
0 = - \partial_{y} [ (- 2 \bar c \,  {\rm sgn}(y) + v_0 
(\sigma_1-\sigma_2) ) P_{\sigma_1,\sigma_2} ] - 2 \gamma P_{\sigma_1,\sigma_2} + \gamma (P_{-\sigma_1,\sigma_2} + P_{\sigma_1,-\sigma_2}) \;.
\ee
Since, up to the sign of $y$ the equation is
linear with constant coefficients, it is natural to look for exponential solutions. In addition, as discussed below Eq. (\ref{Langeviny}), we anticipate also the presence of a delta-function
at $y=0$. This leads us to look for a solution of the form  
\be \label{ansatz} 
P_{\sigma_1,\sigma_2}(y) = A^\epsilon_{\sigma_1,\sigma_2} e^{- \mu |y|} + B_{\sigma_1,\sigma_2} \delta(y) \;,
\ee
where $\mu > 0$ (to be fixed later) and $\epsilon = {\rm sgn}(y)$. For a given $\sigma_1, \sigma_2$, there is no reason a priori, that the solution $P_{\sigma_1, \sigma_2}(y)$ is symmetric around $y=0$, even though
the potential $V(y) = \bar{c} |y|$ is symmetric around $y=0$. This is because the dynamics of $y$ also depends explicitly on $\sigma_1$ and $\sigma_2$, and not just on $y$ alone.   
Hence we put different sets of constants in front of the exponentials in (\ref{ansatz}) for $y>0$ and $y<0$ and they are denoted by different vectors $A^+_{\sigma_1, \sigma_2}$ and 
 $A^-_{\sigma_1, \sigma_2}$. Note that each of them is a 4-component column vector, hence we have $8$ different unknown constants. However, they are related via the
 symmetry relations $A^+_{\sigma_1,\sigma_2} = A^-_{\sigma_2,\sigma_1}$. This follows from the fact that Eq. (\ref{FP}) is invariant under the simultaneous change 
$y \to -y$ and $(\sigma_1,\sigma_2) \to (\sigma_2,\sigma_1)$. Hence, it suffices to know for instance just the vector $A^+$, which has thus four unknown constants. In addition, the amplitudes of the delta-function defined in Eq. (\ref{ansatz}) also form a $4$ component column vector, with $4$ additional unknown constants. Therefore, in total, we have $8$ constants to determine.

By analyzing Eq. \eqref{FPS} around $y=0$, we arrive at two types of conditions. The first one is
that upon injecting the form \eqref{ansatz} in \eqref{FPS} there should be no term generated proportional
to $\delta'(y)$ which implies that for any $\sigma_1,\sigma_2$
\be \label{cond1} 
(\sigma_1-\sigma_2) B_{\sigma_1,\sigma_2} = 0 \;.
\ee
As a consequence we obtain $B_{+-}=B_{-+}=0$. In addition, due to the symmetry $y \to -y$ and $(\sigma_1, \sigma_2) \to (-\sigma_1, - \sigma_2)$ we expect that $B_{++}=B_{--}$. Summarising
\bea \label{B_relation}
B_{+-}=B_{-+}=0 \quad, \quad B_{++}=B_{--} \;.
\eea 
Hence, for the vector $B$, we have only one unknown constant to determine. Combining $A^{\epsilon}$ (with $\epsilon =\pm 1$) and $B$, we then have a total of 5 unknown constants to determine. 
Hence we need 5 relations to fix them. One of them is provided by the normalisation condition, namely $\int_{-\infty}^\infty P(y)\, dy = 1$. The rest of the four conditions can be derived by integrating the FP equations (\ref{FPS}) over a small region across $y=0$. This reads
\be \label{bc}
[ (2 \bar c \,  {\rm sgn}(y) - v_0 
(\sigma_1-\sigma_2) ) P_{\sigma_1,\sigma_2} ]^{0^+}_{0^-} - 2 \gamma B_{\sigma_1,\sigma_2} + \gamma (B_{-\sigma_1,\sigma_2} + B_{\sigma_1,-\sigma_2}) =0 \;,
\ee 
where the second term comes from the contribution of the delta function in \eqref{ansatz}.
Evaluating the first term gives
\be \label{first_term}
[ (2 \bar c \,  {\rm sgn}(y) - v_0 
(\sigma_1-\sigma_2) ) P_{\sigma_1,\sigma_2} ]^{0^+}_{0^-}  = 2 \bar c (A^+_{\sigma_1,\sigma_2} + A^-_{\sigma_1,\sigma_2})
- v_0 (\sigma_1-\sigma_2) (A^+_{\sigma_1,\sigma_2} - A^-_{\sigma_1,\sigma_2}) \;.
\ee
Substituting \eqref{first_term} in \eqref{bc} gives us the four required conditions, namely
\be \label{cond0} 
2 \bar c (A^+_{\sigma_1,\sigma_2} + A^-_{\sigma_1,\sigma_2})
- v_0 (\sigma_1-\sigma_2) (A^+_{\sigma_1,\sigma_2} - A^-_{\sigma_1,\sigma_2}) 
- 2 \gamma B_{\sigma_1,\sigma_2} + \gamma (B_{-\sigma_1,\sigma_2} + B_{\sigma_1,-\sigma_2}) =0  \;,
\ee
for $\sigma_1 = \pm 1$ and $\sigma_2 = \pm 1$. These four conditions (\ref{cond0}) in addition to the normalisation condition provide us exactly 5 relations to determine the 5 unknown constants. In addition, we need to determine the value of $\mu$, to which we now turn to. 

To determine $\mu$, we insert \eqref{ansatz} in \eqref{FPS} in the stationary state and find that 
the amplitude vector $A^{\epsilon}$ must satisfy
\bea \label{MA} 
{\cal M}_\epsilon(\mu) \cdot \left(
\begin{array}{c}
 A^\epsilon_{++} \\
 A^\epsilon_{+-}  \\
 A^\epsilon_{-+}  \\
 A^\epsilon_{--}  \\
\end{array}
\right)
= 0 \;,
\eea 
where we have defined the $4 \times 4$ matrices
${\cal M}_\pm(\mu)$ as
\bea \label{defM}
{\cal M}_\epsilon(\mu) = (- 2 \mu \bar c   - 2 \gamma) \, \mathbb{I} + M(\epsilon \mu) \quad , \quad M(\mu) = \left(
\begin{array}{cccc}
 0  & \gamma  & \gamma  & 0 \\
 \gamma  &  2 \mu  v_0 & 0 & \gamma  \\
 \gamma  & 0 & - 2  \mu  v_0  & \gamma  \\
 0 & \gamma  & \gamma  & 0  \\
\end{array}
\right) \;.
\eea 
The relations in Eq. (\ref{MA}) provide a set of 4 linear equations for the $A^\epsilon_{\sigma_1, \sigma_2}$. The solutions for the $A^\epsilon_{\sigma_1, \sigma_2}$ are identically zero, unless the determinant of ${\cal M}_\epsilon(\mu)$ vanishes. This condition that $\det {\cal M}_\epsilon(\mu) = 0$ actually fixes the value of $\mu$. 
To compute the determinant, we need to evalute the eigenvalues of ${\cal M}_\epsilon(\mu)$, which thanks to Eq. (\ref{defM}), amounts to computing the eigenvalues of the matrix $M(\mu)$. They are given by
\be
\left(0, 0, -2  \sqrt{\gamma^2 + \mu^2 v_0^2}, \;2  \sqrt{\gamma^2 + \mu^2 v_0^2}\right) \;,
\ee 
and the associated eigenvectors are given by the columns of the $4 \times 4$ matrix $\hat O$ 
\bea \label{vp} 
\hat O = \left(
\begin{array}{cccc}
 - \frac{1}{\sqrt{2}}  & \frac{c}{\sqrt{2}} & \frac{s}{2} & \frac{s}{2} \\
 0 & \frac{s}{\sqrt{2}}  & - \frac{1}{2} (1+ c) & \frac{1}{2} (1-c)  \\
 0 & -\frac{s}{\sqrt{2}}  & - \frac{1}{2} (1- c)  & \frac{1}{2} (1+c)  \\
 \frac{1}{\sqrt{2}} & \frac{c}{\sqrt{2}}  & \frac{s}{2}  & \frac{s}{2}  \\
\end{array}
\right) \quad , \quad c = \frac{- \mu v_0}{\sqrt{\gamma^2 + \mu^2 v_0^2} }
\quad , \quad s = \frac{\gamma}{\sqrt{\gamma^2 + \mu^2 v_0^2}} \;,
\eea
with $c^2 + s^2=1$. We will denote the four column vectors respectively by  $(V^1,V^2(\mu),V^3(\mu),V^4(\mu))$. Each of the 
$V^\alpha$ with $\alpha = 1, 2, 3, 4$ is a 4-column vector and they form an orthonormal basis. Hence we get
\bea \label{detM}
\det {\cal M}_\epsilon(\mu) = \left(- 2 \mu \bar c   - 2 \gamma\right) ^2 \left(- 2 \mu \bar c   - 2 \gamma -2  \sqrt{\gamma^2 + \mu^2 v_0^2}\right)\left(- 2 \mu \bar c   - 2 \gamma +2  \sqrt{\gamma^2 + \mu^2 v_0^2}\right) = 0 \;.
\eea
We can obtain different solutions for $\mu$ by setting each of the factors (corresponding to four different eigenvalues of ${\cal M}_\epsilon(\mu)$) to zero. However, it turns out that only the last eigenvalue (corresponding to the last factor in Eq. (\ref{detM})) gives a real positive solution for $\mu$ which reads  
\be
 \label{mu}
\mu = \mu^* = \frac{2 \bar c \gamma}{v_0^2 - \bar c^2} \;,
\ee
where we recall that we are studying the case $v_0>\bar c$, such that there is a bound state. The solution for $A^{\epsilon}$, corresponding to this fourth eigenvalue is therefore $A^{\epsilon} \propto V^4(\epsilon \mu)$, i.e.,
\be \label{AV4}
A^\epsilon_{\sigma_1,\sigma_2} = a V^4_{\sigma_1,\sigma_2}(\epsilon \mu^*) = a \left(
\begin{array}{c}
 \frac{s}{2} \\
 \\
 \frac{1}{2} (1- \epsilon c)  \\
 \\
 \frac{1}{2} (1+ \epsilon c) \\
 \\
 \frac{s}{2}  \\
\end{array}\right) \quad , \quad c = \frac{- \mu^* v_0}{\sqrt{\gamma^2 + (\mu^*)^2 v_0^2} } = \frac{- 2 v_0 \bar c}{v_0^2 + \bar c^2}
~,~ s = \frac{\gamma}{\sqrt{\gamma^2 + (\mu^*)^2 v_0^2}} =  \frac{v_0^2- \bar c^2}{v_0^2 + \bar c^2} \;,
\ee
where $a$ is an a priori unknown amplitude determined below, and $c$ and $s$ are given in \eqref{vp}
with $\mu^*$ given in \eqref{mu}. Note that the symmetry $A^\epsilon_{\sigma_1,\sigma_2} = A^{-\epsilon}_{\sigma_2,\sigma_1}$ discussed above implies that $a$ does
not depend on $\epsilon$. This is because, under this symmetry, the eigenvector $V^4_{\sigma_1,\sigma_2}(\epsilon \mu^*)$ in Eq. (\ref{AV4}) remains invariant, hence $a$ cannot depend
on $\epsilon$. Thus we have reduced the problem of determining 4 unknown constants in the column-vector $A^+_{\sigma_1, \sigma_2}$ to the problem of determining just one constant $a$. Thus to summarize, at this stage, we have two unknowns $a$ and $B_{++}$ to determine.

\begin{figure}[t]
    \centering
    \includegraphics[width = 0.5 \linewidth]{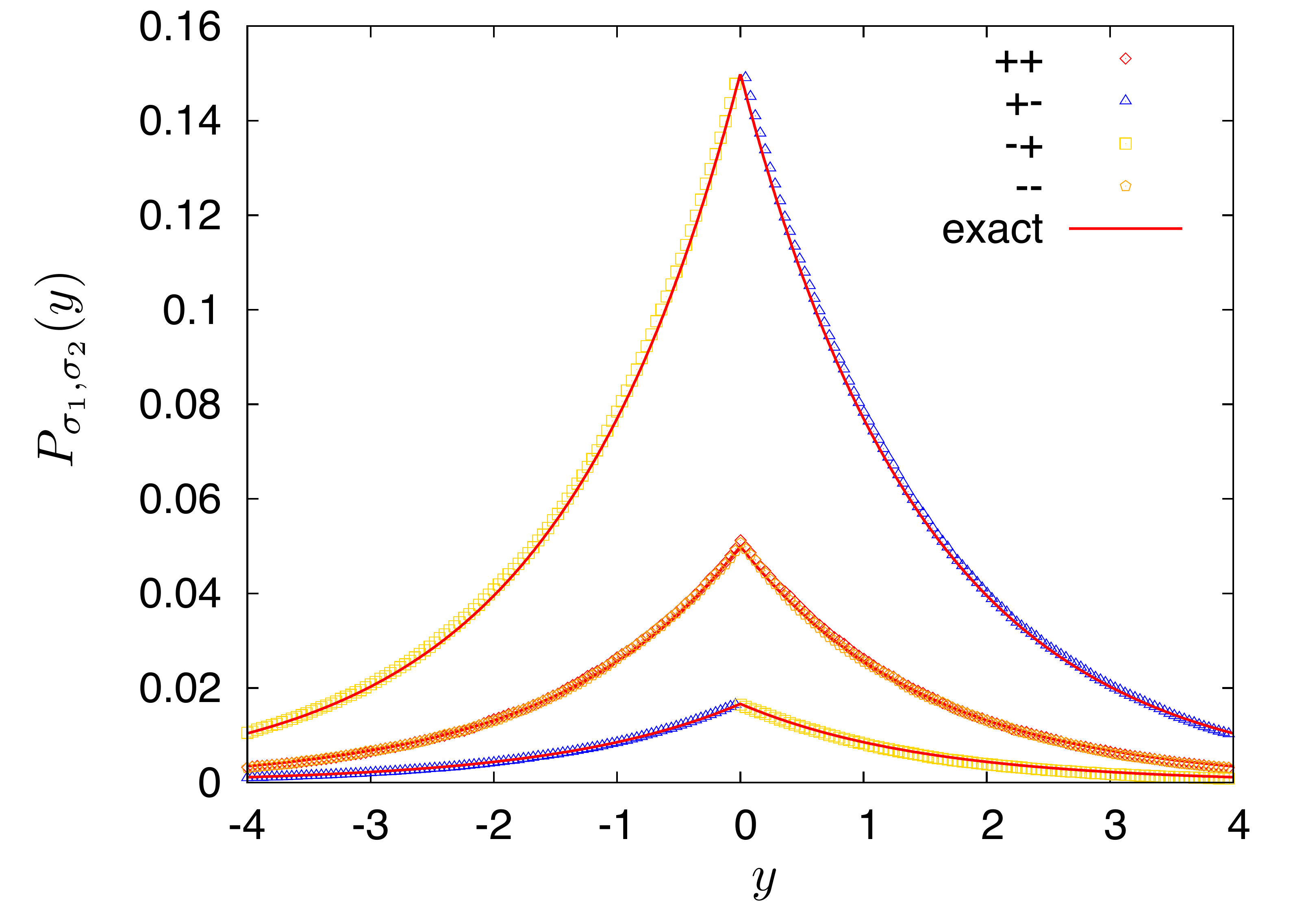}
    \caption{Plot of $P_{\sigma_1, \sigma_2}(y)$ vs $y$ for $v_0 = 1$, $\bar {c} = 1/2$, $\gamma = 1/2$ and $D=0$. The symbols correspond to numerical simulations for $\sigma_1 = \pm 1, \sigma_2 = \pm 1$, obtained by solving Eq. (\ref{Langeviny}), while the solid line corresponds to the exact result in Eq. (\ref{result2}). Note that $P_{++}(y)=P_{--}(y)$ and $P_{+-}(y)=P_{-+}(-y)$ as a consequences of the unicity of the stationary state and of the
    symmetries discussed below in \eqref{sym1}. Note that $P_{+-}(y)$ as well as $P_{-+}(y)$ are both discontinuous at $y=0$, in agreement with (\ref{result2}). Instead, $P_{++}(y) = P_{--}(y)$ exhibit a Dirac delta component $\propto \delta(y)$ which, for clarity, is not shown on the figure, although it is clearly seen on the simulation and its weight is in full agreement with the prediction in (\ref{result2}).}
    \label{fig:zeroD}
\end{figure}
\begin{figure}[ht]
    \centering
    \includegraphics[width = 0.5\linewidth]{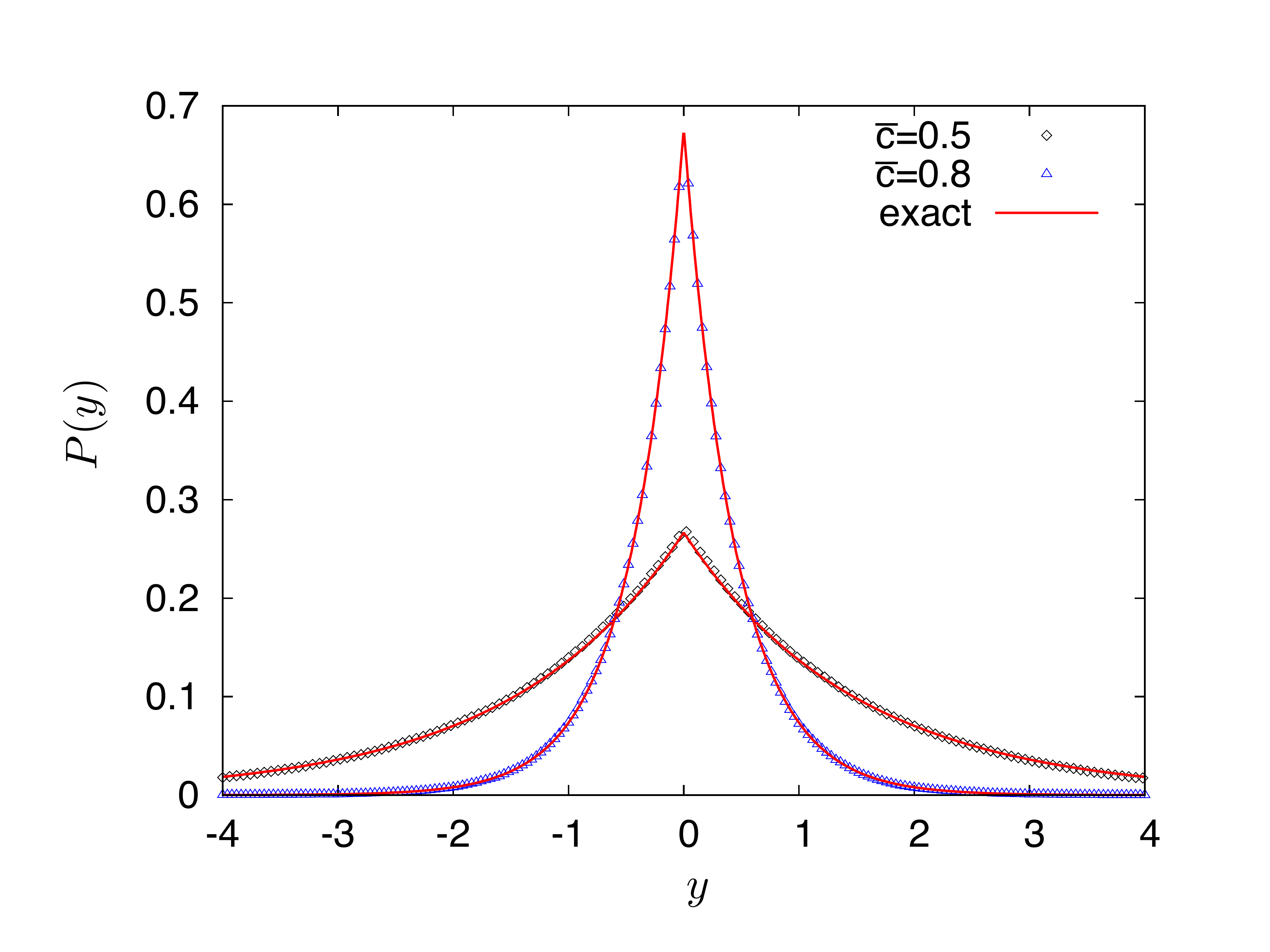}
    \caption{Plot of the total probability density $P(y)$ vs $y$ for two different values of the interaction strength $\bar c=0.5$ and $\bar c=0.8$ for $v_0=1$, $\gamma=1/2$ and $D=0$. The symbols are the results of numerical simulations, obtained by solving Eq. (\ref{Langeviny}), while the solid lines correspond to the exact result in Eq. (\ref{tot_prob}). Note that, for clarity, the Dirac delta component $\propto \delta(y)$ is not shown here although it is clearly seen in the simulations and its weight is in full agreement with the prediction in (\ref{tot_prob}).}
    \label{fig:totzeroD}
\end{figure}

To proceed, we first rewrite the condition \eqref{cond0} explicitly 
\be \label{cond3} 
2 \bar c a
\left(
\begin{array}{c}
 s \\
 1  \\
 1  \\
 s  \\
\end{array}
\right) + 
v_0 a
\left(
\begin{array}{c}
 0 \\
 2 c  \\
 2 c  \\
 0  \\
\end{array}
\right) + 
\left(
\begin{array}{cccc}
 -2 \gamma  & \gamma  & \gamma  & 0 \\
 \gamma  &  -2 \gamma & 0 & \gamma  \\
 \gamma  & 0 & -2 \gamma  & \gamma  \\
 0 & \gamma  & \gamma  & - 2 \gamma  \\
\end{array}
\right) \, 
\left(
\begin{array}{c}
 B_{++} \\
 B_{+-}  \\
 B_{-+}  \\
 B_{--}  \\
\end{array}
\right) = 0 \;.
\ee
Next, we use Eq. (\ref{B_relation}) to eliminate $B_{+-}, B_{-+}$ and $B_{--}$ in favour of $B_{++}$. This gives the two relations 
\bea
&& 2 a \bar c s - 2 \gamma B_{++} = 0 \;, \\
&& 2 a (\bar c  + v_0 c ) + 2 \gamma B_{++} = 0 \;,
\eea
which are actually equivalent using the values for $c$ and $s$ from Eq. (\ref{AV4}). This leads to the single relation
\be
B_{++}= a \, \bar c \frac{v_0^2- \bar c^2}{\gamma(v_0^2 + \bar c^2)} \;.
\ee
We are then left with one unkwown constant $a$ to determine and this will be fixed by the normalization condition. Injecting these results in the form \eqref{ansatz} and summing over $\sigma_1,\sigma_2$ we obtain the total
probability
\be
P(y) = \sum_{\sigma_1 = \pm 1, \sigma_2 = \pm 1} P_{\sigma_1, \sigma_2}(y) = a (s+1) e^{- \mu^* |y|} + 2 B_{++} \delta(y) = 
2 a \left( \frac{v_0^2}{v_0^2 + \bar c^2} e^{- \mu |y|} + \bar c \frac{v_0^2 - \bar c^2}{\gamma(v_0^2 + \bar c^2)} \delta(y)\right) \;.
\ee
Imposing the normalization condition
$\int_{-\infty}^{+\infty} dy P(y) =1$
then allows to determine $a$ as
\be
a = \frac{1}{2} \frac{\bar c \gamma}{v_0^2- \bar c^2} \;,
\ee
which leads to the final explicit result for the stationary probability $P_{\sigma_1,\sigma_2}(y)$
\bea \label{result2} 
 P_{\sigma_1,\sigma_2}(y) =  \frac{\gamma \bar c}{4(v_0^2 + \bar c^2)}\;  
 e^{- \frac{2 \gamma \bar c}{v_0^2 - \bar c^2} |y|} 
 \left(
\begin{array}{c}
 1  \\
\frac{v_0+\bar c}{v_0-\bar c}  \theta(y) + 
\frac{v_0-\bar c}{v_0+\bar c}  \theta(-y) \\
\frac{v_0-\bar c}{v_0+\bar c}   \theta(y) + \frac{v_0+\bar c}{v_0-\bar c}  \theta(-y)  \\
1  \\
\end{array}
\right)  + \frac{1}{2} \frac{\bar c^2}{v_0^2 + \bar c^2}
 \delta(y) \left(
\begin{array}{c}
 1 \\
0 \\
0 \\
1  \\
\end{array} \right)  \;,
\eea
as well as the total probability
\be \label{tot_prob}
P(y) = \frac{\bar c \gamma v_0^2}{v_0^4 - \bar c^4}\; e^{- \frac{2 \gamma \bar c}{v_0^2 - \bar c^2} |y|} + \frac{\bar c^2}{v_0^2 + \bar c^2}
 \delta(y) \;.
\ee
In Fig. \ref{fig:zeroD} we compare our theoretical results for $P_{\sigma_1, \sigma_2}(y)$ in Eq. (\ref{result2}) for $v_0 = 1$, $\bar {c} = 1/2$, $\gamma = 1/2$ and $D=0$ with numerical simulations, showing a perfect agreement (note that, to keep the figure readable, the Dirac delta components of $P_{++}(y)$ and $P_{--}(y)$ are not shown there although they are clearly seen on the simulations -- see also Fig. \ref{fig:traj} -- and we have checked that their associated weight fully agrees with the prediction in (\ref{result2})). In Fig. \ref{fig:totzeroD} we compare our result for the total probability $P(y)$ in Eq. (\ref{tot_prob}) for two different values of $c=0.5$ and $c=0.8$ (and $v_0=1$, $\gamma=1/2$ and $D=0$) with numerical simulations, showing also a very good agreement. 

Note also that for each state $(\sigma_1,\sigma_2)$ one can check from Eq. \eqref{result2} that 
$\int_{-\infty}^{+\infty} dy P_{\sigma_1,\sigma_2}(y) = \frac{1}{4}$, hence each of the four states is equiprobable in the stationary solution, 
as expected. The variance of the position is
\be
\int_{-\infty}^{+\infty} dy \, y^2 \, P(y) = \frac{v_0^2}{2 \gamma^2 \bar c^2} \frac{(v_0^2-\bar c^2)^2}{v_0^2 + \bar c^2} \;.
\ee 
In the passive limit $\gamma, v_0 \to +\infty$ with $D_{\rm eff} = v_0^2/(2 \gamma)$, as well as $\bar c$, fixed, the weight of the delta function
vanishes as $\bar c^2/v_0^2$ and one recovers the standard Gibbs-Boltzmann distribution
\be
P(y ) \to \frac{\bar c}{2 D_{\rm eff}} e^{- \bar c |y|/D_{\rm eff}} \quad , \quad P_{\sigma_1,\sigma_2}(y) \to \frac{1}{4} \frac{\bar c}{2 D_{\rm eff}} e^{- \bar c |y|/D_{\rm eff}}  \;,
\ee
for all $\sigma_1,\sigma_2$. Therefore the delta-peak in Eq. (\ref{tot_prob}) in the stationary distribution is an explicit signature of the activity in the system. Note that even for finite $v_0$, the non-delta function part of $P(y)$ in \eqref{tot_prob} retains a pure symmetric-exponential form $\propto e^{-\mu |y|}$, as in the passive case, albeit with a different decay rate $\mu = 2\gamma \bar c/(v_0^2-\bar{c}^2)$ from the passive case $\propto e^{- \bar c|y|/D_{\rm eff}}$. 

When $v_0 \to \bar c^+$ each term in $P(y)$ in \eqref{tot_prob} goes to $\frac{1}{2} \delta(y)$ and the size of the bound state goes to zero. To investigate the fine structure inside the 
critical regime, one can rescale $y$ by the typical size of the bound state. Denoting $y= \frac{v_0-\bar c}{\gamma} \tilde y $ one obtains the scaling form in the critical region as
\be \label{res_simple}
P(y) dy \to \tilde P(\tilde y) d\tilde y \quad , \quad \tilde P(\tilde y) =   \frac{1}{4} e^{- |\tilde y|} \left(
\begin{array}{c}
 0 \\
1 \\
1 \\
0  \\
\end{array} \right)   + \frac{1}{2} \delta(\tilde y) \left(
\begin{array}{c}
 1 \\
0 \\
0 \\
1  \\
\end{array} \right) \;.
\ee 
The first part shows that when they have opposite velocities $\sigma_1=-\sigma_2$ the
two RTP's form a (very small) exponential bound state (weak clustering), while when they have identical velocities they
are bound at exactly the same position in space (strong clustering).

\subsection{With thermal noise, $D>0$} \label{sec:linDnon0}


We now switch on a nonzero value of $D$ in Eq. (\ref{relative_D0}). 
%
As a result, the FP equation for $P_{\sigma_1,\sigma_2}(y,t)$ changes from Eq.~(\ref{FP}) to
\bea \label{FPD}
\partial_t P_{\sigma_1,\sigma_2} = - \partial_{y} [ (- 2 \bar c \,  {\rm sgn}(y) + v_0 
(\sigma_1-\sigma_2) ) P_{\sigma_1,\sigma_2} ] - 2 \gamma P_{\sigma_1,\sigma_2} + \gamma (P_{-\sigma_1,\sigma_2} + P_{\sigma_1,-\sigma_2})
+ 2 D \partial_{y}^2 P_{\sigma_1,\sigma_2} \;,
\eea 
where only the last term on the r.h.s., involving the second derivative with respect to $y$, is $D$-dependent. We now look for a stationary solution, setting $\partial_t P_{\sigma_1,\sigma_2}=0$ in the l.h.s of \eqref{FPD}.
Since $D>0$ this solution will obey:

(i) continuity of $P_{\sigma_1,\sigma_2}(y)$ at $y=0$

(ii) a jump in the derivative at zero, with the matching condition
\be \label{match} 
P'_{\sigma_1,\sigma_2}(0^+)-P'_{\sigma_1,\sigma_2}(0^-) = - 2 \frac{\bar c}{D} 
P_{\sigma_1,\sigma_2}(0) \;.
\ee 
These give two sets of four conditions, since they hold for any $\sigma_1,\sigma_2$. As discussed in Section \ref{sec:model}, we anticipate that the presence of a finite $D$ will smear out the delta function and replace it by a cusp at $y=0$ and exponential decaying profile, whose width will vanish as $D \to 0^+$. Since, up to the sign of $y$, Eq. (\ref{FPD}) is
linear with constant coefficients, it will be a linear superposition of exponentials for $y>0$ and $y<0$ separately.
We thus look for a particular solution of the form
\be \label{ansatzD} 
P_{\sigma_1,\sigma_2}(y) = A^\epsilon_{\sigma_1,\sigma_2} e^{- \mu |y|} 
\ee
where $\epsilon = {\rm sgn}(y)$.

Let us start by determining $\mu$. Inserting \eqref{ansatzD} in \eqref{FPD} in the stationary state we find that 
the amplitude vector $A^{\epsilon}$ must satisfy the same condition \eqref{MA} where now the matrix ${\cal M}_\pm(\mu)$ 
has an additional $D$-dependent diagonal term 
\bea
{\cal M}_\epsilon(\mu) = (- 2 \mu \bar c + 2 D \mu^2    - 2 \gamma) \, \mathbb{I} + M(\epsilon \mu) 
\eea
with $M(\mu)$ given in \eqref{defM}. The eigenvalues of ${\cal M}_\epsilon(\mu)$ and their associated eigenvectors
are then
\bea
&& - 2 \mu \bar c + 2 D \mu^2 - 2 \gamma  \quad , \quad V^1 \\
&& - 2 \mu \bar c + 2 D \mu^2 - 2 \gamma  \quad , \quad V^2(\epsilon \mu) \\
&& - 2 \mu \bar c + 2 D \mu^2 - 2 \gamma -2  \sqrt{\gamma^2 + \mu^2 v_0^2} \quad , \quad V^3(\epsilon \mu)  \\
&& - 2 \mu \bar c + 2 D \mu^2 - 2 \gamma + 2 \sqrt{\gamma^2 + \mu^2 v_0^2} \quad , \quad V^4(\epsilon \mu)
\eea 
where the eigenvectors $\hat O= (V^1,V^2(\mu),V^3(\mu),V^4(\mu))$ are given (in column form) in \eqref{vp} and depend on $\mu$ via $\bar {c}$ and~$s$.
As in the previous sub-section, the value of $\mu$ is fixed by the condition 
\bea \label{detM_D}
\det {\cal M}_\epsilon(\mu) &=& \left(- 2 \mu \bar c + 2 D \mu^2 - 2 \gamma \right)^2 \left(  - 2 \mu \bar c + 2 D \mu^2 - 2 \gamma -2  \sqrt{\gamma^2 + \mu^2 v_0^2} \right) 
\left( - 2 \mu \bar c + 2 D \mu^2 - 2 \gamma + 2 \sqrt{\gamma^2 + \mu^2 v_0^2} \right) \nonumber \\
&=& 0 \;.
\eea

\vspace*{0.5cm}
\noindent
{\bf First two eigenvectors.} Setting the first factor in (\ref{detM_D}) to $0$ (corresponding to the sectors of the two first eigenvectors $V^1$ and $V^2$) we get
\be
- 2 \mu \bar c  + 2 D \mu^2 - 2 \gamma = 0 \;.
\ee 
Taking the positive root we obtain
\be \label{mu12}
\mu = \mu_1= \mu_{2}=  \frac{\bar c  + \sqrt{\bar c^2 + 4 D \gamma}}{2 D} \;,
\ee 
which corresponds to a component of $P_{\sigma_1,\sigma_2}(y)$ proportional to
$e^{- \frac{\bar c + \sqrt{ \bar c^2 + 4 D \gamma}}{2 D} |y| }$. In the limit $D \to0$
this component yields the $\delta(y)$ term obtained in the previous section for $D=0$ 
[see Eq. (\ref{result2})]. 

\vspace*{0.5cm}
\noindent
{\bf The two other eigenvectors}. The third and the fourth factors in Eq. (\ref{detM_D}) (corresponding to the eigenvectors $V^3$ and $V^4$) lead to the pair of equations
\bea \label{eqn2} 
- \mu \bar c  + D \mu^2 - \gamma +   \nu \sqrt{\gamma^2 + \mu^2 v_0^2}= 0 \;,
\eea 
with $\nu=-1$ for $V^3$ and $\nu=+1$ for $V^4$. For $D=0$ one finds the solutions $\mu=0$ and $\mu=  \frac{2 \bar c \gamma}{v_0^2 - \gamma^2}$ found previously. For $D>0$, 
let us use dimensionless units. We write $\mu = \frac{\gamma}{v_0} \tilde \mu$ and look for the {positive} roots $\tilde \mu$ of
\bea \label{eqn22} 
f_\eta(\tilde \mu)= - g \tilde \mu   + \tilde D \tilde \mu^2 - 1 +   \nu \sqrt{1 + \tilde \mu^2}= 0 \;,
\eea 
in terms of the two dimensionless parameters
\be
g=\frac{\bar c}{v_0} \quad , \quad \tilde D = \frac{D \gamma}{v_0^2} \;.
\ee
Taking the square of Eq. (\ref{eqn22}) one finds a quartic equation for $\tilde \mu$. However, one can easily check that there is a trivial solution $\tilde \mu = 0$, which is
of course discarded since we need $\tilde \mu >0$. This leads to a cubic equation for $\tilde \mu$
\be \label{cubic1}
-2 g   + \tilde \mu  \left(2   \tilde D+ 1- g^2\right)
+2 g \tilde D \tilde \mu ^2  - \tilde D^2 \tilde \mu ^3   = 0 \;.
\ee
Note that after squaring Eq. (\ref{eqn22}) the $\nu$-dependence has disappeared. Thus the information about the associated eigenvenctor ($V^3$ or $V^4$) has been lost. 
Hence, we need to reinject the solution for $\tilde \mu$ [from Eq. (\ref{cubic1})] back into the original unsquared Eq. (\ref{eqn22}) to recover the eigenvector dependence. Indeed, by noting
that $\sqrt{1 + \tilde \mu^2}>0$ and $\nu = \pm 1$, it follows from Eq. (\ref{eqn22}) that
\bea \label{rel_nu}
\nu = {\rm sgn}\left( 1 + g \tilde \mu + \tilde D \tilde \mu^2 \right) \;,
\eea
with $\nu = -1$ associated to $V^3$ while $\nu = +1$ associated to $V^4$.


To solve the cubic equation (\ref{cubic1}) we first rewrite it in the standard form, by  writing $\tilde \mu = t + {2 g}/{(3 \tilde D)}$, which gives 
$t^3 + p t + q = 0$ with
\be
  p = -\frac{6 \tilde D+g^2+3}{3 \tilde D^2} \quad , \quad q = \frac{2 g \left(9 \tilde D+g^2-9\right)}{27 \tilde D^3} \;,
\ee
and the discriminant is given by 
\be
\Delta_2 = - (4 p^3 + 27 q^2) = 4 \tilde D^{-6} \left((\tilde D^2 +10 \tilde D-2) g^2+(2 \tilde D+1)^3+g^4\right) \;.
\ee
It is easy to see that $\Delta_2$ is always positive, hence there are 3 real roots indexed by $k=0,1,2$ and given by the Cardano's formulae \cite{cardano_wiki}
\be \label{tk}
t_k = 2 \sqrt{ - \frac{p}{3} } \cos\left( \frac{1}{3} \arccos\left(\frac{3 q}{2 p} \sqrt{- \frac{3}{p}}\right) - \frac{2 \pi k}{3} \right) \quad, \quad k = 0, 1, 2  \;.
\ee
This leads to three possible roots in the original variable $\tilde \mu$
\be \label{mu01} 
\tilde \mu_k = \frac{2 g}{3 \tilde D} + \frac{2}{3} \sqrt{\frac{6 \tilde D+g^2+3}{\tilde D^2}} \cos
   \left(\frac{1}{3} \cos ^{-1}\left(-\frac{g \left(9
   \tilde D+g^2-9\right)}{\left(6
   \tilde D+g^2+3\right)^{3/2}}\right) - \frac{2 \pi k}{3} \right) \;.
\ee
By investigating Eq. (\ref{mu01}) using Mathematica, we find that  
\bea
&& \tilde \mu_0 > \tilde \mu_1 > 0 \quad , \quad \tilde \mu_2 <0 \;.
\eea
Since $\tilde \mu_2 < 0$ we discard this root. Thus the only allowed roots are $\tilde \mu_0$ and $\tilde \mu_1$. 
Now using Eq. (\ref{rel_nu}), we find that the values of $\nu$ associated to these two roots are respectively $\nu = -1$ for $\tilde \mu_0$ and $\nu = +1$ for $\tilde \mu_1$. Thus $\tilde \mu_0$ is associated to the eigenvector $V^3$ while $\tilde \mu_1$ is associated to $V^4$. Hence in summary the only positive roots are 
\bea \label{mu34} 
\mu_3 = \frac{\gamma}{v_0} \tilde \mu_0 \quad , \quad V^3 \\
\mu_4 = \frac{\gamma}{v_0} \tilde \mu_1 \quad , \quad V^4 \;. \nonumber 
\eea
\\

\begin{figure}[t]
    \centering
    \includegraphics[width = 0.5 \linewidth]{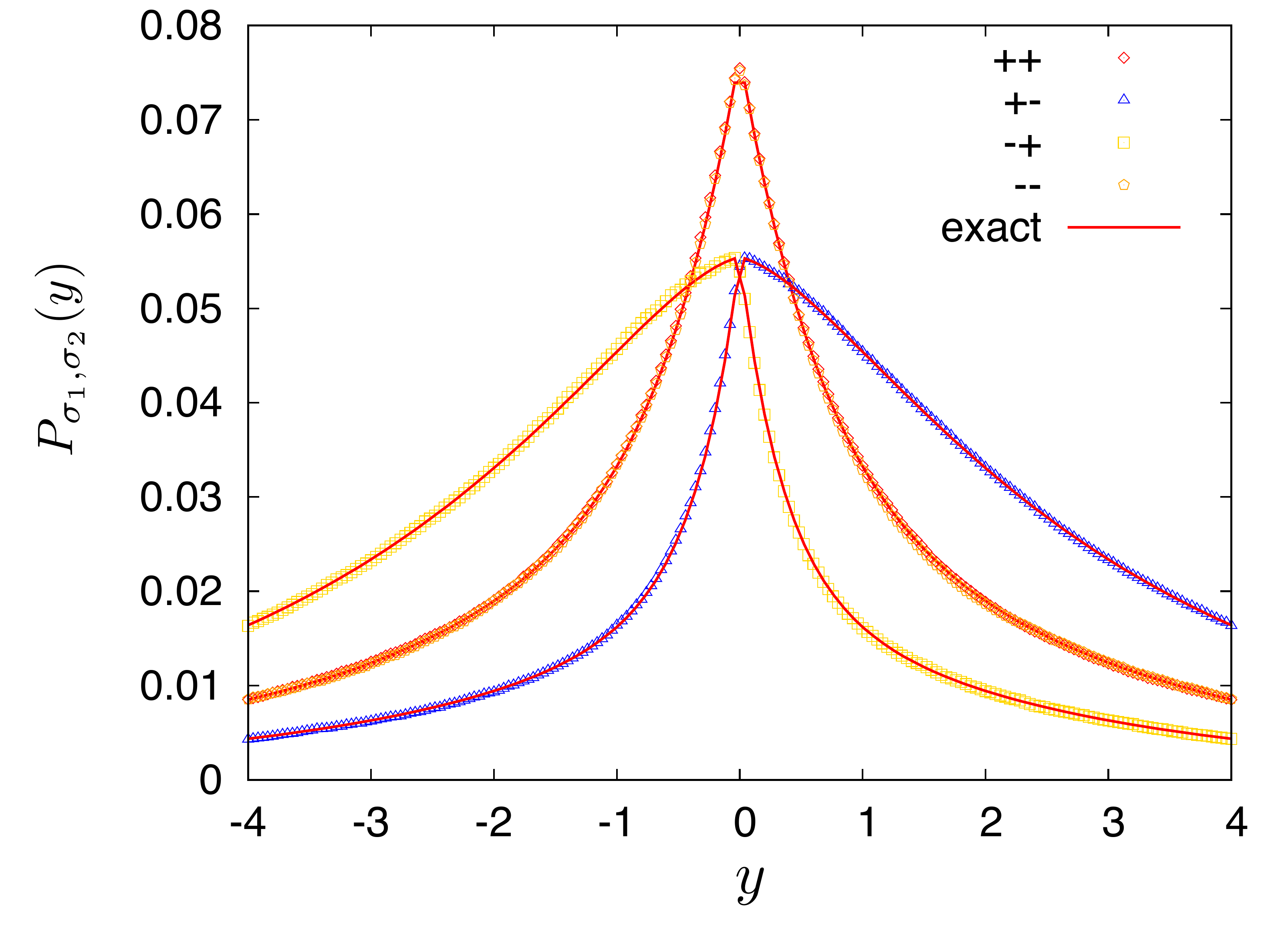}
    \caption{Plot of $P_{\sigma_1, \sigma_2}(y)$ vs $y$ for $v_0 = 1$, $\bar {c} = 1/2$, $\gamma = 1/2$ and $D=1/2$. The symbols correspond to numerical simulations for $\sigma_1 = \pm 1, \sigma_2 = \pm 1$ while the solid line corresponds to the exact result in Eq. (\ref{statnew2}). Note that $P_{++}(y)=P_{--}(y)$ $P_{+-}(y)=P_{-+}(-y)$ as a consequences of the unicity of the stationary state and of the
    symmetries discussed below in \eqref{sym1}.}
    \label{fig:finiteD}
\end{figure}
To conclude the general solution is
\bea \label{full_P}
&& P_{\sigma_1,\sigma_2}(y) = [(b_1 V^1_{\sigma_1,\sigma_2} + b_2 
V^2_{\sigma_1,\sigma_2}(\mu_2) )    
\theta(y) +
(b'_1 V^1_{\sigma_1,\sigma_2} + b'_2 
V^2_{\sigma_1,\sigma_2}(- \mu_2) ) \theta(-y) ]
e^{- \mu_{2} |y| }  \\
&& + (b_3 V^3_{\sigma_1,\sigma_2}(\mu_3) \theta(y) + 
b'_3 V^3_{\sigma_1,\sigma_2}(- \mu_3) \theta(-y) ) e^{- \mu_3 |y|} 
+ (b_4 V^4_{\sigma_1,\sigma_2}(\mu_4) \theta(y) + 
b'_4 V^4_{\sigma_1,\sigma_2}(- \mu_4) \theta(-y) ) e^{- \mu_4 |y|} \;. \nonumber 
\eea
To determine the 8 coefficients $b_i,b'_i$ we can use 
(i) the continuity of $P_{\sigma_1,\sigma_2}(y)$ at $y=0$ (4 equations) (ii) the 4 matching conditions \eqref{match} 
\be \label{match2} 
P'_{\sigma_1,\sigma_2}(0^+)-P'_{\sigma_1,\sigma_2}(0^-) = - 2 \frac{\bar c}{D} 
P_{\sigma_1,\sigma_2}(0) \;,
\ee 
and then the normalization condition for the total probability $P(y)=\sum_{\sigma_1=\pm 1} \sum_{\sigma_2=\pm 1} P_{\sigma_1,\sigma_2}(y)$,
i.e., $\int_{-\infty}^{+\infty} dy P(y)=1$. This is performed in the Appendix, leading to the result given in Eq. (\ref{statnew2}). In Fig. \ref{fig:finiteD}, we compare this analytical predictions for $P_{\sigma_1, \sigma_2}(y)$ in Eq. (\ref{statnew2}) for a specific set of the parameters of the model to numerical simulations, showing an excellent agreement.

To conclude this section, we provide our explicit results for the final result for the total probability $P(y)$ 
\bea \label{fullP_Dpos}
&& P(y) = \frac{1}{2(\frac{c_2}{\mu_2} + \tilde b_3 \frac{s_3-1}{\mu_3} + 
\tilde b_4 \frac{s_4+1}{\mu_4} )} \left(
c_2 e^{- \mu_{2} |y| }   + \tilde b_3 (s_3-1) e^{- \mu_3 |y|} 
+ \tilde b_4 (s_4+1) e^{-\mu_4 |y|} \right) \\
&& \tilde b_3 = \frac{s_2 \left(\bar c-D \mu
   _4\right)}{\left(c_3+c_4\right) \bar c - D \left(c_4 \mu _3+c_3 \mu
   _4\right)} \quad , \quad \tilde b_4=  \frac{s_2 \left(\bar c-D \mu
   _3\right)}{\left(c_3+c_4\right) \bar c - D \left(c_4 \mu _3+c_3 \mu
   _4\right)}
\eea 
and we recall that
\bea
c_i = - \frac{\mu_i v_0}{\sqrt{\gamma^2 + \mu_i^2 v_0^2} }
\quad , \quad s_i = \frac{\gamma}{\sqrt{\gamma^2 + \mu_i^2 v_0^2}} 
\eea 
where $\mu_2$ is given in \eqref{mu12}, $\mu_3$ and $\mu_4$ are given in \eqref{mu34}-\eqref{mu01}.
\begin{figure}
    \centering
    \includegraphics[width = 0.5\linewidth]{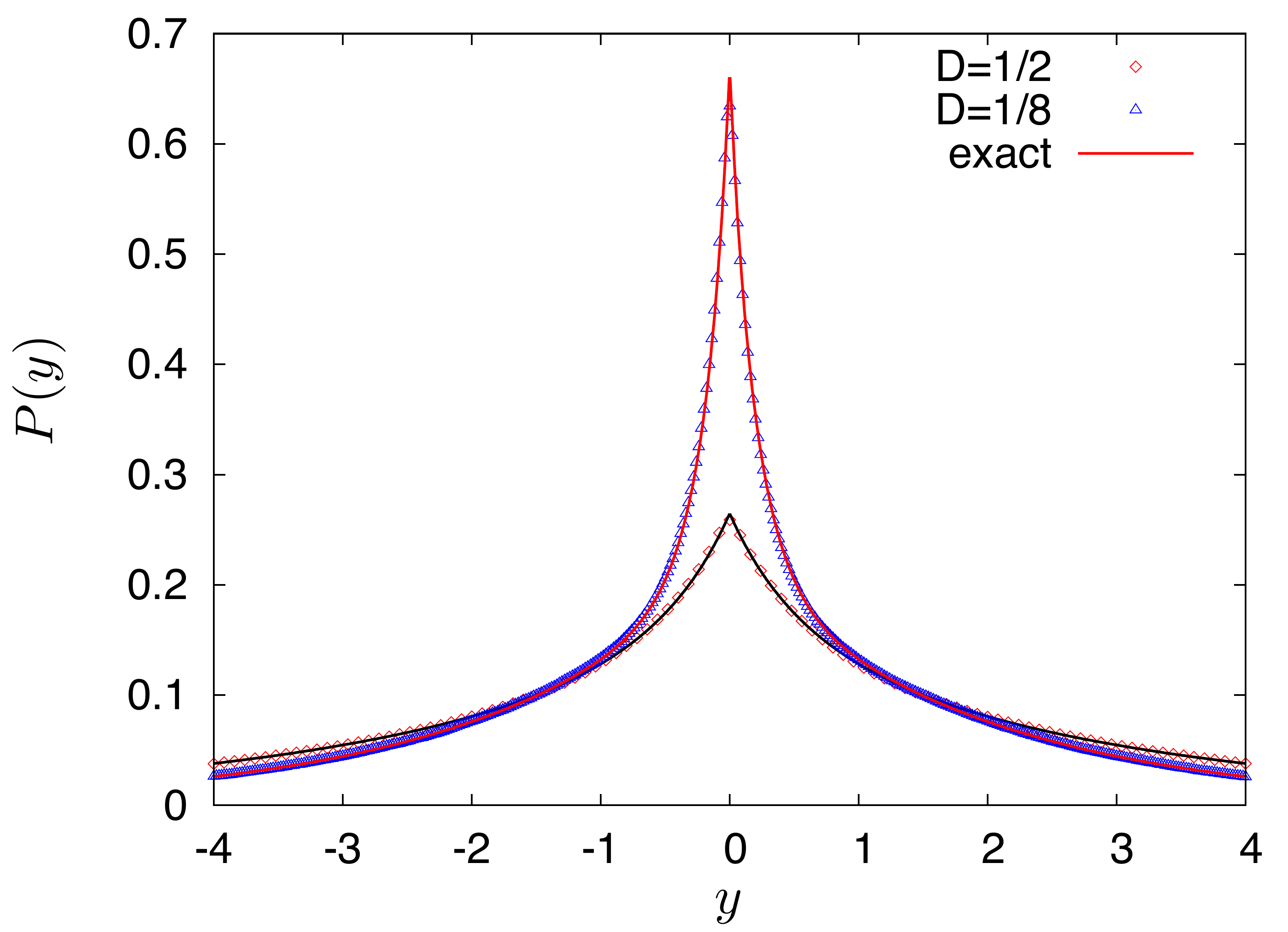}
    \caption{Plot of the total probability density $P(y)$ vs $y$ for two different values the diffusion constant $D=1/2$ and $D=1/8$ and for $\bar c=0.5$, $v_0=1$, $\gamma=1/2$. The symbols are the results of numerical simulations while the solid lines correspond to the exact results in Eq. (\ref{exact_D2}) for $D=1/2$ and in Eq. (\ref{exact_D8}) for $D=1/8$.}
    \label{fig:totnonzeroD}
\end{figure}
For instance for $\bar c=1/2$, $D=1/2$, $v_0=1$, $\gamma=1/2$ we find
\be
P(y)=
{{0.020441}}\, e^{-3.3234 y}+ {{0.0868356}} \, e^{-1.61803 y}+ {{0.157553}} \, e^{-0.357926 y} \;, \label{exact_D2}
\ee
while for $\bar c=1/2$, $D=1/8$, $v_0=1$, $\gamma=1/2$ 
we find
\be
P(y)= {{0.0523053}} \, e^{-12.331 y}+{{0.397104}} \, e^{-4.82843 y}+{{0.220603}}\, e^{-0.533482 y} \;. \label{exact_D8}
\ee 
These theoretical predictions in Eqs. (\ref{exact_D2}) and (\ref{exact_D8}) are compared to numerical simulations in Fig. \ref{fig:totnonzeroD}, showing a very good agreement.


\section{More general interaction without diffusion} \label{sec:gen}

\subsection{Models and flow diagram} 

In this section we discuss the case of a more general attractive interaction between the two RTP's, i.e., a more general force $f(y)$ which is an odd
function of $y$, $f(-y)=-f(y)$. For simplicity, we will set $D=0$ as argued earlier. In this case, the equations for the center of mass $w=(x_1+x_2)/2$
and the relative coordinate $y=x_1-x_2$ respectively in (\ref{cofm}) and (\ref{relative0}) reduce to 
\begin{eqnarray} \label{relative} 
\frac{dw}{dt}&=&  \frac{v_0}{2}\, (\sigma_1(t) + \sigma_2(t))  \;, \label{cofm1}  \\
\quad \frac{dy}{dt}&=& 2 f(y) + v_0\, (\sigma_1(t) - \sigma_2(t) )  \;. \label{relative2}
\end{eqnarray}
%
%
In principle, we can write down the FP equation for the joint distribution $P_{\sigma_1, \sigma_2}(w,y,t)$. This joint distribution obviously does not reach a steady
state, since $w(t)$ in Eq. (\ref{cofm1}) corresponds to a free RTP motion and hence diffuses at late times. Only the marginal distribution of the relative coordinate 
$P_{\sigma_1,\sigma_2}(y,t) = \int_{- \infty}^{+\infty} dw \, P_{\sigma_1,\sigma_2}(w,y,t)$ reaches a steady state as $t \to \infty$. Hence, we focus on the $y$-marginal only.
The FP equation for $P_{\sigma_1,\sigma_2}(y,t)$ reads
\bea \label{2_fp.3}
\partial_t P_{\sigma_1,\sigma_2} = - \partial_{y} [ (2f(y) + v_0 
(\sigma_1-\sigma_2) ) P_{\sigma_1,\sigma_2} ] - 2 \gamma P_{\sigma_1,\sigma_2} + \gamma (P_{-\sigma_1,\sigma_2} + P_{\sigma_1,-\sigma_2}) \;.
\eea 
Note that for $f(y) = - \bar c \,{\rm sgn}(y)$ discussed in Section \ref{sec:lin}, this equation reduces to Eq. (\ref{FP}). 
\\
\begin{figure}[t]
\centering
\includegraphics[width=0.6\linewidth]{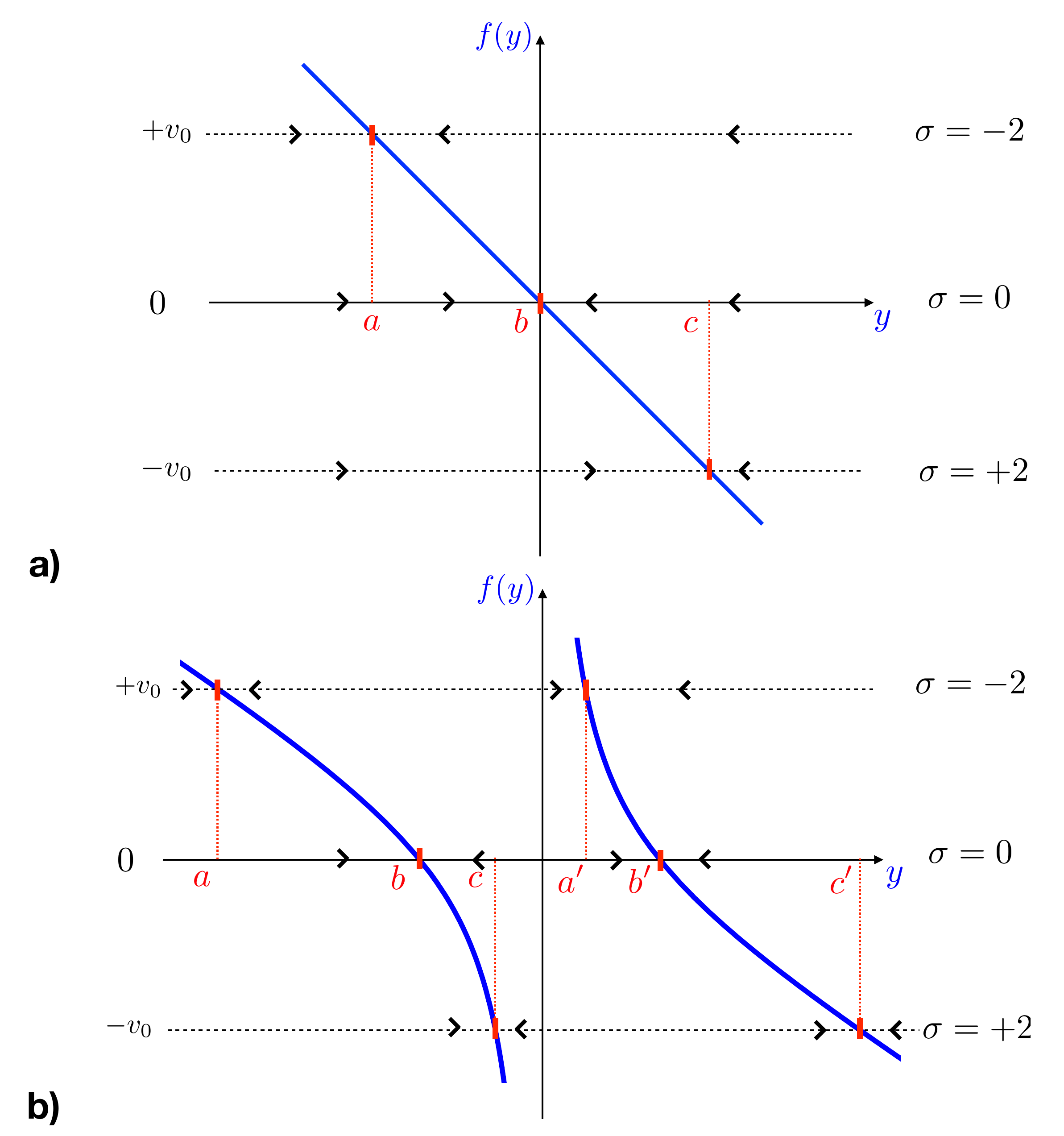}
\caption{Dynamical diagrams for the 2 RTP's within the two models discussed in the text. The interaction force between the RTP's, $f(y)$
is plotted versus $y$, the relative coordinate. The fixed points for the dynamics (\ref{relative2}) in each state of the two RTP's
are determined by the roots of the equations $f(y)=- v_0$ (state $+-$), $f(y)=0$ (states $++$ and $--$)
and $f(y)=v_0$ (state $-+$). These are obtained as the intersections of the plot of $f(y)$ with the three horizontal lines which correspond to the three possible values of $\sigma = \sigma_1 - \sigma_2$. {\bf Top panel a)}: the harmonic force model, $f(y)=-y$, with the three fixed points $a, b$ and $c$ (see
discussion in the text). The flow indicated by the arrows shows the dynamics of $y(t)$ in each state. The three fixed
points are attractive and the steady state is supported by the interval $[a,c]$. 
{\bf Bottom panel b)}: the second model (with a repulsive short range interaction), $f(y)=\frac{1}{y}-y$, with the six fixed points $a, b,c$ and $a',b',c'$ (see
discussion in the text). All fixed points are attractive. The steady state is not unique, with two
supports $[a',c']$ and $[a,c]$, depending on the initial ordering of the particles.}
\label{fig:Cot}
\end{figure}
To search for a stationary solution $P_{\sigma_1,\sigma_2}(y)$, one must set the l.h.s. of Eq. (\ref{2_fp.3}) to $0$. 
However, before writing the steady state equations, it is useful to see what we can anticipate about the form of this solution, and in particular
whether this solution is unique. 
This can be guessed by studying the stability behaviour of the Langevin equation~(\ref{relative2}), following the examples in Refs. \cite{DKM19,LMS2020}. 
We will consider two generic examples of interactions between the RTP's, which present a different steady state behavior. 
The first example is the harmonic force $f(y) = - \lambda y$. In the second example there is in addition a repulsion between the particles
so that they cannot cross, with $f(y)= \frac{1}{y} - \lambda y$. The curves $f(y)$ versus $y$ are plotted in 
Fig. \ref{fig:Cot} for both examples.
Note that $\sigma_1(t) - \sigma_2(t)$
can take only three values $-2\,v_0,0,2\, v_0$ corresponding respectively to the two RTP's being in the states $(-+)$, $(++)$ or $(--)$ and $(+-)$.  
Hence to find stationary points for a fixed 2 particle state one must look at the roots of the equations, respectively $f(y)=v_0$, $f(y)=0$ and $f(y)=- v_0$. 
In Fig. \ref{fig:Cot} we have indicated the positions of these roots. In the first example (harmonic force) there are three of them denoted $a$, $b$ and $c$.
Because of the symmetry $f(-y)=-f(y)$ one has $b=0$ and $c=-a=\lambda/v_0$. In the second example (with short range repulsion) there are six of them: 
three on the $y<0$ side (i.e., $x_1<x_2$) denoted $a$, $b$ and $c$ and three on the $y>0$ positive side (i.e., $x_1>x_2$), denoted $a'$, $b'$ and $c'$.
Because of the symmetry $f(-y)=-f(y)$ one has $a'=-c$, $b'=-b$ and $c'=-a$. We have indicated by arrows in the figure the flow diagram for each of the three values of $\sigma=\sigma_1(t) - \sigma_2(t)$. For both examples considered here all fixed points $a,b,c$ (and $a',b',c'$) are attractive (i.e., $f'(y)$ is negative at the fixed point). 
From the general analysis performed in Ref. \cite{LMS2020}, we can predict the support of the steady state probabilities from the flow in the Fig. \ref{fig:Cot}.

\vspace*{0.5cm}
\noindent{\bf In the first model (harmonic force)} we predict that the stationary state is unique and that after a finite time 
the relative coordinate $y(t)$ will end up within the interval $[a,c]$. Once the particle enters this interval $[a,c]$, it can never go out via the dynamics in Eq. (\ref{relative2}). Hence we would expect that the stationary state, if it exists, will be supported only inside the interval $[a,c]$. In other words, the stationary probability density $P_{\sigma_1,\sigma_2}(y)$ will strictly vanish outside this
interval $[a,c]$.

\vspace*{0.5cm}
\noindent{\bf In the second model (with short range repulsion)} the two RTP's cannot cross, hence it is clear that, depending on the initial condition, i.e., whether 
$x_1(0)> x_2(0)$ or $x_2(0)> x_1(0)$, the relative coordinate $y(t)$ of the two RTP's will end up either on $[a,c]$ or on $[a',c']$. In that case the 
stationary solution is not unique, and there are two possible disconnected supports, which are images of each other by the symmetry $y \to -y$. 

\vspace*{0.5cm}

\subsection{Determination of the steady state solution} 

We now focus on the case where the stationary solution is unique (for models in the same class as the harmonic well) and show how to
find this solution. Let us rewrite the equation Eq. (\ref{2_fp.3}) in components and set $\partial_t P_{\sigma_1,\sigma_2}(y,t) =0$,
leading to the four coupled equations 
\bea 
&& \partial_t P_{++} = - 2 \partial_y (f(y) P_{++}) 
 - 2 \gamma P_{++}  
+ \gamma (P_{+-}+P_{-+} )= 0 \label{2_fp.31} \\
&& \partial_t P_{+-} = - 2 v_0 \partial_y P_{+-}  - 2 \partial_y (f(y) P_{+-})  
 - 2 \gamma P_{+-}  
+ \gamma (P_{++}+P_{--} ) = 0 \label{2_fp.32} \\
&& \partial_t P_{-+} = 2 v_0 \partial_y P_{-+}  - 2 \partial_y (f(y) P_{-+}) 
  - 2 \gamma P_{-+}  
+ \gamma (P_{++}+P_{--} ) = 0 \label{2_fp.33} \\
&&  \partial_t P_{--} =   - 2 \partial_y (f(y) P_{--})  
 - 2 \gamma P_{--}  
+ \gamma (P_{+-}+P_{-+} ) = 0 \;. \label{2_fp.34}
\eea
These equations (see also \eqref{2_fp.3}) are invariant under the change $(y,\sigma_1,\sigma_2) \to (y,-\sigma_2,-\sigma_1)$. Since
$f(y)$ is an odd function of $y$, they are also invariant under the change $(y,\sigma_1,\sigma_2) \to (-y,-\sigma_1,-\sigma_2)$.
Since we consider here the case where the stationary solution is unique, this implies that 
\bea \label{sym1}
P_{\sigma_1, \sigma_2}(y) = P_{-\sigma_1, -\sigma_2}(-y) \quad , \quad P_{\sigma_1, \sigma_2}(y) = P_{\sigma_2, \sigma_1}(-y) \;,
\eea
where the second symmetry is obtained by combining the two symmetries mentioned above. The first corresponds to reversing the speed of each particle and reversing the direction of $y$. 
Since the confining potential $V(y)$ is symmetric under $y \to - y$ (equivalently $f(-y)=-f(y)$), the first symmetry in Eq. (\ref{sym1}) is evident
when the stationary state is unique. Summing over $\sigma_1,\sigma_2$ this also implies that 
the total probability $P(y)$ must be an even function of $y$.

It is convenient to introduce the following quantities
\bea
p_1 = P_{++} + P_{- -} \quad , \quad 
p_2= P_{+ +} - P_{- -} \quad , \quad 
q_1= P_{+ -} + P_{- +} \quad , \quad 
q_2= P_{+ -}- P_{- +} \quad , \quad P = p_1+q_1 \;.
\eea 
In terms of these quantities Eqs. (\ref{2_fp.31})-(\ref{2_fp.34}) simplify to
\bea
&& -\partial_y [ f(y) p_1]-\gamma p_1 + \gamma q_1=0  \label{1} \\
&& -\partial_y [f(y) p_2] - \gamma p_2=0  \label{2} \\
&& -\partial_y[f(y) q_1 + v_0 q_2]  - \gamma q_1+ \gamma p_1=0  \label{3} \\
&& -\partial_y[ f(y) q_2 + v_0 q_1]  - \gamma q_2=0 \label{4}  \;.
\eea 
Amazingly, Eq. (\ref{2}) for $p_2(y)$ completely decouples from $p_1, q_1$ and $q_2$ for arbitrary $f(y)$. In fact $p_2$
is not needed to obtain $p_1,q_1$ and $q_2$. The origin of this simplification will be discussed below.
It therefore remains to solve the three equations (\ref{1}), (\ref{3}) and (\ref{4}). Adding \eqref{1} and \eqref{3} and using $P = p_1 + q_1$ one finds
\be \label{current}
\partial_y ( 2f(y) P(y) + 2v_0 q_2(y)) = 0 \quad \Longrightarrow \quad
2f(y) P(y) +2 v_0 q_2(y) = J \;,
\ee
where $J$ is a constant. We can identify this constant with the total probability current in the system. This can be seen by adding the four equations (\ref{2_fp.3}) for $\sigma_1 = \pm 1$ and
$\sigma_2 = \pm 1$ which gives $\partial_t P = - \partial_y J(y)$ with $J(y)= 2f(y) P(y) + 2v_0 q_2(y)$. In the steady state, the probability current must be a constant since $\partial_t P = 0$. Hence $J(y) = J$ is independent of $y$ and coincides with Eq. (\ref{current}).  Since $f(y)$ is an odd function, $P(y)$ is even, and $q_2(y)$ is also an odd function because of the symmetry \eqref{sym1} and of the unicity of the steady state, the equation \eqref{current}
implies that the current $J$ must vanish. 
Setting $J=0$ we get another relation
\be \label{current2}
 f(y) P(y)=f(y) (p_1(y) + q_1(y)) =- v_0 q_2(y) \;.
\ee 
We now eliminate $q_2(y)$ from Eqs. \eqref{3} and \eqref{4} by using the relation in Eq. (\ref{current2}). This gives a pair of coupled equations, involving $P(y)$ and $q_1(y)$
\bea
&& - \partial_y (f(y) (q_1-P)) + \gamma P -2 \gamma q_1 = 0 \label{a1} \\
&& - \partial_y (v_0^2 q_1 - f(y)^2 P) + \gamma f(y) P = 0 \label{a2} \;,
\eea 
which can be conveniently re-written as
\bea
&& f P' + (f'+ \gamma) P = f q_1' + (f'+2 \gamma) q_1 \label{a11}\\
&& f^2 P' + f(2 f' + \gamma) P = v_0^2 q_1'  \;. \label{a12}
\eea 
After differentiating both equations (\ref{a11}) and (\ref{a12}) and performing straightforward manipulations, one can eliminate $q_1$ and write a closed second order ordinary differential equation for $P(y)$. We get
\bea \label{final} 
&& f(y)
   \left(v_0^2-f(y)^2\right) P''(y) +  \left(\left(v_0^2-3 f(y)^2\right) \left(\gamma +2 f'(y)\right)+\frac{f(y)
   \left(f(y)-v_0\right) \left(f(y)+v_0\right) f''(y)}{2 \gamma +f'(y)}\right) P'(y) \\
   && +
   \left(\frac{\gamma  \left(v_0^2-3 f(y)^2\right) f''(y)}{2 \gamma +f'(y)}-f(y)
   \left(\gamma +2 f'(y)\right) \left(2 \gamma +3 f'(y)\right)\right)P(y)= 0\;.
\eea
One can in principle solve this equation for $P(y)$ using the boundary conditions given below. Once $P(y)$ is known one obtains 
$q_2(y)= - f(y) P(y)/v_0$ from \eqref{current2}. One also obtains
$q_1(y)$ by integration of \eqref{a12}. Alternatively, a similar second order differential equation can also be derived for $q_1(y)$ by eliminating $P(y)$ from the pair of Eqs. (\ref{a11}) and (\ref{a12}). We do not write it explicitly here.  
As argued before, the stationary solution is expected to be supported over the finite interval $[a,c]$ where $f(a) = v_0$ and $f(c) = -v_0$. Therefore Eq. (\ref{final}) for $P(y)$
holds for $y \in [a,c]$. In addition, we need to provide the appropriate boundary conditions to find the unique solution. These  boundary conditions are nontrivial and we derive them below. 

\vspace*{0.5cm}
\noindent{\bf Boundary conditions.} The main idea is to derive, directly from the Langevin equation (\ref{relative2}) how the four probabilities $P_{\sigma_1, \sigma_2}(y,t)$ evolve in a small time $\Delta t$ exactly at the two edges of the support $y=a$ and $y=c$. Let us illustrate this explicitly with the state $P_{++}(y,t)$. For this case, the Langevin equation (\ref{relative2}) says that in a small time $\Delta t$ the position of the particle evolves by $\Delta y = 2\,f(y) \, \Delta t$. Therefore the evolution of the probability density $P_{++}(y,t)$ can be written as
\bea \label{FP_deltat}
P_{++}(y,t+\Delta t) = (1-2 \gamma \Delta t)  P_{++}(y-2\,f(y) \, \Delta t,t) +  \gamma \Delta t \left[P_{+-}(y,t) + P_{-+}(y,t)  \right] \;.
\eea
This is easily explained since in the time interval $[t, t+\Delta t]$ the velocities $(v_0 \sigma_1(t), v_0 \sigma_2(t))$ do not change sign with probability $1 - 2 \gamma \Delta t$. Hence, if the particle wants to be at the location $y$ at time $t+ \Delta t$, it must have been at $y - \Delta y = y-2\,f(y) \, \Delta t$ at time $t$. This explains the first term in Eq. (\ref{FP_deltat}). In contrast, with probability 
$\gamma \Delta t$, it can come from the state $(+-)$ or $(-+)$ just by flipping the negative velocity. This explains the last two terms in Eq.~(\ref{FP_deltat}). Now we consider this evolution equation (\ref{FP_deltat}) exactly at the left edge $y=a$ where we recall that $f(a) = v_0$. Hence the first term on the r.h.s. of Eq. (\ref{FP_deltat}) reads $(1-2 \gamma \Delta t)P_{++}(a-2 v_0 \Delta t,t)$. Since $2 v_0 \Delta t > 0$, the argument $a-2 v_0 \Delta t < a$. This means that the argument $a-2 v_0 \Delta t$ is outside the left edge of the support where, by definition, there is no particle in the stationary state. Hence the first term is identically zero at $y=a$. As $\Delta t \to 0$, the last two terms in Eq. (\ref{FP_deltat}) also vanish. This gives us the boundary condition in the stationary state
\bea \label{bc1}
P_{++}(y=a) = 0  \;.
\eea
By repeating this argument for each of the states $(\sigma_1 = \pm 1, \sigma_2 = \pm 1)$ at the two boundaries $a$ and $c$, we find the following set of boundary conditions
\bea 
&&P_{++}(a) = P_{++}(c) = 0 \label{bc_++} \\
&&P_{--}(a) = P_{--}(c) = 0 \label{bc_--} \\
&& P_{+-}(a) = 0 \label{bc_+-} \\
&&P_{-+}(c) = 0 \label{bc_-+} \;. 
\eea
Note that, due to the symmetry condition (\ref{sym1}), these boundary conditions are not all independent. In fact, there are only four independent boundary conditions. Since our original stationary states equations (\ref{2_fp.31})-(\ref{2_fp.34}) are four first-order differential equations (albeit coupled), these four boundary conditions are enough to fix the stationary solution uniquely. 
Note that these boundary conditions \eqref{bc_++}-\eqref{bc_-+} mean that no jump is allowed at these points for these probabilities, which must thus
vanish continuously. 

Let us now return to the function $p_2(y)= P_{++}(y) - P_{--}(y)$. Using both symmetries in Eq. (\ref{sym1}) for $\sigma_1 = \sigma_2 = +$
we obtain that 
\bea \label{argument1}
p_2(y) = P_{++}(y) - P_{--}(y) = P_{++}(y) - P_{++}(-y) = 0 \;.
\eea
Hence all the components $P_{\sigma_1,\sigma_2}(y)$ of the stationary state can be determined.
\\

\subsection{Harmonic interactions and mapping to a three state model} 

\begin{figure}
    \centering
    \includegraphics[width = 0.5\linewidth]{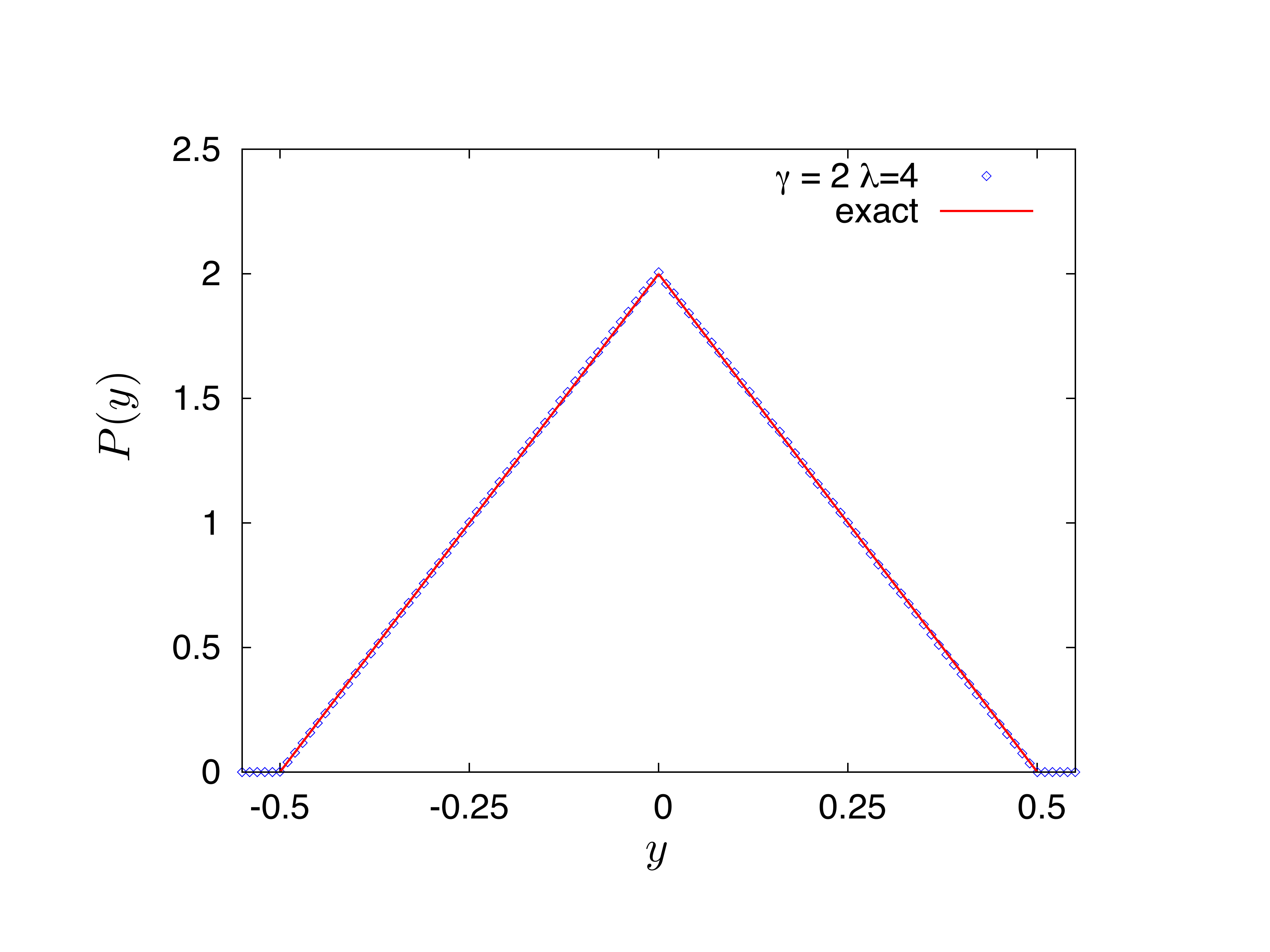}
    \caption{Plot of the total probability density $P(y)$ vs $y$ for the harmonic interaction $f(y) = - \lambda y$ with $v_0=1$, $\lambda = 2$ and $\gamma=2 \lambda = 4$. The symbols are the results of numerical simulations while the solid lines correspond to the exact result in Eq.~(\ref{exact_quad}).}
    \label{fig:totnonzeroD_harmo}
\end{figure}
To illustrate the method, let us consider the example of the harmonic interaction $f(y)=-\lambda y$ in which case
Eq.~\eqref{final} becomes
\bea \label{eqharm} 
(\gamma -2 \lambda ) P'(y) \left(v_0^2-3 \lambda ^2 y^2\right)-\lambda  y P''(y)
   \left(v_0^2-\lambda ^2 y^2\right)+\lambda  y (2 \gamma -3 \lambda ) (\gamma -2 \lambda )
   P(y) = 0 \;.
\eea
Let us first study the special case $\gamma=2 \lambda$, which turns out to be a bit simpler. In that case, indeed, the above equations
simplify into 
\bea
&& y \left(v_0^2-\lambda ^2 y^2\right) P''(y) = 0 \\
&& - \lambda y P' + \lambda P = - \lambda y q_1' + 3 \lambda q_1 \\
&& \lambda^2 y^2 P'  = v_0^2 q_1' \;.
\eea 
The solution is easily obtained for $y \in [- \frac{v_0}{\lambda}, \frac{v_0}{\lambda}]$ as 
\be  \label{exact_quad}
P(y)=\frac{\lambda^2}{v_0^2} \left(\frac{v_0}{\lambda} - |y|\right)  \quad , \quad  q_1(y)= \frac{1}{3} \frac{\lambda}{v_0} - \frac{\lambda^4}{3 v_0^4} {\rm sgn}(y) y^3 \\
\quad , \quad  q_2(y)= \frac{\lambda^3}{v_0^3} y \left(\frac{v_0}{\lambda} - |y|\right) \;,
\ee 
and zero for $|y|>\frac{v_0}{\lambda}$. In Fig. \ref{fig:totnonzeroD_harmo} we show a comparison of our exact prediction in Eq. (\ref{exact_quad}) with numerical simulations for $\gamma = 2 \lambda = 4$ (as well as $v_0=1.0$) showing a very good agreement.

The general solution of \eqref{eqharm} can be obtained in terms of hypergeometric functions. In fact, as we show below, the present
problem can be mapped onto a recently studied problem of a single RTP with position $y$, with three internal states, in an harmonic potential $V(y)=\frac{1}{2} \lambda y^2$,
which was solved in terms of hypergeometric functions \cite{3statesBasu}. This leads to the general result for the solution of the 2 RTP problem with harmonic interaction, i.e. of Eq. \eqref{eqharm},
as 

\bea
P(y) &=& A_1~\left [ _2F_1\left(1- \frac \beta 2,\frac 32 - \beta,\frac{3-\beta}2;\left(\frac{\lambda y}{v_0}\right)^2 \right) \right. \cr
 &&\left. +\frac{2}{\sqrt{\pi}} \frac{\Gamma(\frac{3-\beta}2)\Gamma(\beta+\frac 12)}{(1- 2 \beta)\Gamma(\frac {\beta+1} 2)}\left(\frac{\lambda y}{v_0}\right) ^{\beta -1}~ _2F_1 \left(\frac 12,1-\frac \beta 2,\frac{\beta +1}2;\left(\frac{\lambda x}{v_0}\right)^2\right) \right] \quad, \quad -\frac{v_0}{\lambda} \leq y \leq \frac{v_0}{\lambda}\label{eq:Px_sol} \;,
\eea
where $\beta = \gamma/\lambda$ and the amplitude $A_1$ is given in Eq. (33) of Ref. \cite{3statesBasu}. 

\begin{figure}[t]
    \centering
    \includegraphics[width=0.8\linewidth]{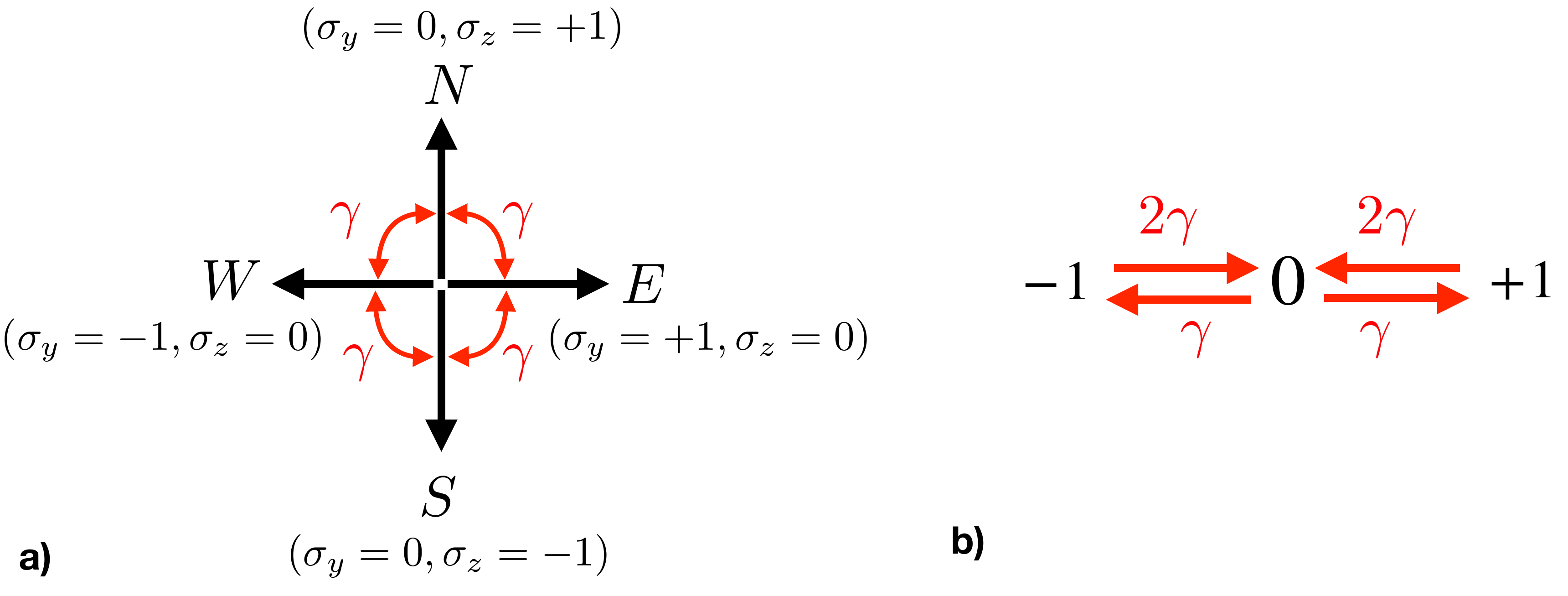}
    \caption{Illustration of the four-state model (left panel) and corresponding three-state model studied in Ref. \cite{3statesBasu} and discussed in the text.}
    \label{fig:3states}
\end{figure}

The model studied in Ref. \cite{3statesBasu} is defined by the transition rates between the three states denoted $+1,0,-1$ as shown in Fig. \ref{fig:3states} b). 
One can identify these states with ours as
\bea 
1 &\equiv& (+ -)  \quad \Rightarrow \quad P_1 = P_{+-} \label{mapping1}\\
-1 &\equiv& (- +) \quad \Rightarrow \quad P_{-1} = P_{-+} \label{mapping-1}\\
0 &\equiv& (+ +) \cup (- -) \quad \Rightarrow \quad P_0 = P_{++}+P_{--} \label{mapping0}\;.
\eea 
This implies that the probabilities denoted as $P$, $Q$ and $R$ in Ref. \cite{3statesBasu} are related to $P$, $q_1$ and $q_2$ studied here via 
\bea
Q = P_{1} + P_{-1} &\equiv& q_1 \;, \\
R = P_{1} - P_{-1} &\equiv& q_2 \;, \\
P = P_0 + P_{1} + P_{-1} &\equiv& P \;.
\eea 
Hence the solution for these functions obtained there also provide the solution for our model. 
\subsection{General mapping to a two-dimensional single RTP model} 

In fact, this mapping, between (i) the relative coordinate of a pair of interacting particles and (ii) the position of a single particle in a confining potential subjected
to a three-state active noise, actually is more general than just the harmonic interaction and can be extended to arbitrary attractive interaction in (i). The general mapping
can be formulated as follows. Consider a single particle on a plane with its coordinates $(y(t), z(t))$ evolving via the pair of equations
%
%
%
\bea
&& \frac{d y(t)}{dt} = f(y(t)) + v_0 \sigma_y(t) \label{Many}\\
&& \frac{d z(t)}{dt} = g(z(t)) + v_0 \sigma_z(t) \label{Manz}
\eea
where the confining force $f(y)$ in the $y$-direction depends only on $y$ and the confining force $g(z)$ in the $z$-direction depends only on $z$. In Eqs. (\ref{Many}) and (\ref{Manz}), $\sigma_{y,z}(t)$ are the $y$ and $z$-components of a director field which has four possible orientations $E$ (East), $N$ (North), $W$ (West) and $S$ (South), as shown in Fig. \ref{fig:3states} a). The transition rates between the different directions of the director field are indicated in Fig. \ref{fig:3states} a).  
The state of the system has four labels $E, N, W$ and $S$ and hence there are four position distribution functions ${\cal P}_E(y,z,t)$, ${\cal P}_N(y,z,t)$, ${\cal P}_W(y,z,t)$ and ${\cal P}_S(y,z,t)$. In Ref. \cite{SBS2020}, these position distribution functions were computed explicitly in the ``free'' case where $f(y)=0$ and $g(z)=0$. Here we consider instead the case of nonzero interaction forces $f(y) \neq 0$ and $g(z) \neq 0$. Taking into account the transition rates in Fig. \ref{fig:3states} a), one can explicitly write down the associated coupled FP equations [along the lines of Eq. (\ref{2_fp.3})]. We do not repeat them here as they are a bit long (see Eqs. (6a)-(6d) of Ref. \cite{3statesBasu} in the case $f(y)=-\mu y$ and $g(z)=-\mu z$). Following Ref. \cite{3statesBasu} we consider three marginal position distribution in the $y$-direction
\bea \label{defPmarg}
&&P_1(y,t) = \int  dz \,~{\cal P}_E(y,z,t) \;, \label{margP1} \\
&&P_{-1}(y,t) = \int dz \,~{\cal P}_W(y,z,t) \;,\label{margPm1} \\
&&P_0(y,t)= \int dz ~[{\cal P}_N(y,z,t) + {\cal P}_S(y,z,t)] \;. \label{margP0}
\eea
From the explicit FP equations for ${\cal P}_{E,N,W,S}$, it is straightforward to obtain the FP equations for these three marginal probability densities proceeding
as in \cite{3statesBasu}
 \bea
 \frac{\partial}{\partial t} P_{1}(y,t) &=& \frac{\partial }{\partial y} [(f(y)- v_0 ) P_{1}] + \frac\gamma 2 P_0 - \gamma P_1 \;, \label{eq:FP_3st1}\\
 \frac{\partial}{\partial t} P_{-1}(y,t) &=& \frac{\partial }{\partial y} [(f(y) + v_0 ) P_{-1}] + \frac\gamma 2 P_0 - \gamma P_{-1} \;, \label{eq:FP_3stm1}\\
 \frac{\partial}{\partial t} P_{0}(y,t) &=& \frac{\partial }{\partial y} [f(y) P_{0}] + \gamma  (P_1 + P_{-1}) - \gamma P_0 \;. \label{eq:FP_3st0}
 \eea 
We first note that the force field $g(z)$ has completely dropped out in Eqs. (\ref{eq:FP_3st1})-(\ref{eq:FP_3st0}) and this is precisely due to the decoupled structure of the force field (\ref{Many}) and (\ref{Manz}). How is this related to the two-particle model with attractive force $f(y)$? If we take our basic FP equations in Eq. (\ref{2_fp.3}) and use the identification in Eqs. (\ref{mapping1}), (\ref{mapping-1}) and (\ref{mapping0}), it is easy to see that the resulting FP equations for $P_1, P_{-1}$ and $P_0$ are exactly the same as in Eqs. (\ref{eq:FP_3st1})-(\ref{eq:FP_3st0}). 
In fact it is possible by comparing Eqs. \eqref{2_fp.31}-\eqref{2_fp.34} in the present paper, and Eqs. (6a)-(6d) of Ref. \cite{3statesBasu}
to establish a direct mapping between the two interacting RTP model and the single RTP $2d$-model \cite{3statesBasu}. One finds
\bea \label{defPmarg}
&& P_{+-}(y,t) = \int  dz \,~{\cal P}_E(y,z,2t) \quad , \quad P_{-+}(y,t) = \int dz \,~{\cal P}_W(y,z,2t) \;,\label{margPm1} \\
&& P_{++}(y,t)= \int dz \,{\cal P}_N(y,z,2t) \quad , \quad P_{--}(y,t) = \int dz \,{\cal P}_S(y,z,2t) \;,\label{margP0}
\eea
the rescaling of time being equivalent to a rescaling of $\gamma$. Note that there is some arbitrariness in connecting $(N,S) \equiv (++,--)$ rather
than $(N,S) \equiv (--,++)$. Note that $p_2(y,t) = P_{++}(y,t) - P_{--}(y,t) =
\int dz \,[{\cal P}_N(y,z,2t) - {\cal P}_S(y,z,2t)]$ indeed decouples as found above since it is determined by the dynamics along $z$ (which is
not observed). This completes the mapping between the $y$-component of a $2d$ single particle problem subjected to a force-field as in Eqs. (\ref{Many})-(\ref{Manz}) and the relative coordinate of a pair of RTP's with arbitrary interactions between them.

Finally, it is interesting to observe that if one instead looks at the marginals of the $z$-coordinate $z$ in the $2d$ model (i.e., if we integrate over $y$ instead of $z$) one has again a mapping
to two RTP's interacting with a force $g(z)$, and the identification $W \equiv --$, $E \equiv ++$, $N \equiv +-$, and $S \equiv -+$
(which is obtained by a rotation of the Fig. \ref{fig:3states} a)).

\section{Conclusion}\label{sec:conclusion}

In this paper we have studied two RTP's interacting via an attractive potential $V(y)$, depending on their relative coordinate $y$. 
In the large time limit, the total probability distribution of $y$ reaches a stationary form $P(y)$ which we have characterized in terms
of the solution of a second-order differential equation. For two specific examples of potential $V(y)$ we have obtained the four components
of the stationary distribution, $P_{\sigma_1,\sigma_2}(y)$ where $\sigma_1=\pm 1$, $\sigma_2=\pm 1$ are the states of each RTP with velocity $\pm v_0$. 

As a first example, we have studied in detail the case of a linear
potential $V(y) \sim |y|$, first without thermal noise (i.e. for $D=0$) and then for general $D>0$. 
In the first case, $D=0$, a striking result is that $P(y)$ is the sum of a delta function part $\delta(y)$ and a decaying exponential.
This is the signature of a strong clustering effect, when the two RTP's are in the same state. This is reminiscent of
the observation in \cite{slowman,slowman2,MBE2019,KunduGap2020}, except that here the clustering effect is enhanced by the attractive interaction.
Because of the finite decay length of the exponential part, the weight of the delta part is finite even for an infinite system.
In the presence of thermal noise, $D>0$, the delta function broadens and $P(y)$ is now a sum of exponentials. We have tested
our analytic formula with numerical simulations of two interacting RTP's. 
 
The second example is the harmonic attraction, $V(y)= \frac{\lambda}{2} y^2$. In that case, the support of $P(y)$ is found to be a finite
interval $[- \frac{v_0}{\lambda} , \frac{v_0}{\lambda}]$. We found that the general solutions for $P(y)$ on this interval are expressed in terms of hypergeometric functions. In addition, $P_{\sigma_1,\sigma_2}(y)$ generically exhibit power-law singular behaviors at the three points $y=0,\pm \frac{v_0}{\lambda}$. Remarkably, 
this exact solution can be mapped onto a problem studied previously of a three-state single RTP in one dimension in a harmonic external potential. 
In fact, as we have shown, this mapping extends to any interaction potential $V(y)$. In addition, a related mapping to a two-dimensional single
RTP model is obtained. Finally, we have also discussed the effect of an additional short-range repulsion. When it is strong enough so that the particles
cannot cross it results in the existence of several distinct steady states which are related by the symmetry~$y \to -y$. 

As we have seen here, it is already non-trivial to obtain the stationary probability for two RTP's with a general interaction potential.
An interesting question is whether there are solvable models for a number $N>2$ RTP's. Preliminary study shows that it is already quite challenging
for the linear attraction potential. Multi-particle models with $N>2$ were studied for a chain of harmonically attracting RTP's, for which 
the mean square displacement of a single RTP as well the two time correlation have been obtained \cite{SinghChain2020,PutBerxVanderzande2019}.
However the stationary distribution has not been studied, and as we see here, already for $N=2$, it involves hypergeometric functions. 
These questions are left for future investigations. 

Recently, the simple RTP model of a single particle $dx/dt= v_0 \sigma(t)$ in Eq. \eqref{def_RTP}
has been generalised to the case $d^nx/dt^n=v_0 \sigma(t)$ with any $n>0$~\cite{DMS21}.
For example, the case $n=2$ would correspond to an undamped particle driven
by a random telegraphic force.  It would be interesting to see if our method can
be extended to study a pair of such interacting undamped RTP's.

\vspace*{0.5cm}
{\it Acknowledgments:} 
This research was supported by ANR grant ANR-17-CE30-0027-01 RaMaTraF.

\appendix

\section{Details for two RTP's with diffusion}

As stated in the text the stationary solution $\partial_t P_{\sigma_1,\sigma_2}(y)=0$ of the FP equation
\eqref{FPD} which obeys (i) continuity at $y=0$ and (ii) the matching condition \eqref{match} for derivatives 
has the form 
\bea
&& P_{\sigma_1,\sigma_2}(y) = [(b_1 V^1_{\sigma_1,\sigma_2} + b_2 
V^2_{\sigma_1,\sigma_2}(\mu_2) )    
\theta(y) +
(b'_1 V^1_{\sigma_1,\sigma_2} + b'_2 
V^2_{\sigma_1,\sigma_2}(- \mu_2) ) \theta(-y) ]
e^{- \mu_{2} |y| }  \\
&& + (b_3 V^3_{\sigma_1,\sigma_2}(\mu_3) \theta(y) + 
b'_3 V^3_{\sigma_1,\sigma_2}(- \mu_3) \theta(-y) ) e^{- \mu_3 |y|} 
+ (b_4 V^4_{\sigma_1,\sigma_2}(\mu_4) \theta(y) + 
b'_4 V^4_{\sigma_1,\sigma_2}(- \mu_4) \theta(-y) ) e^{- \mu_4 |y|} \;, \nonumber 
\eea
where the eigenvectors $\hat O= (V^1,V^2(\mu),V^3(\mu),V^4(\mu))$ are given (in column form) in \eqref{vp} and depend on $\mu$ via $c$ and~$s$.
The parameter $\mu_2$ is given in \eqref{mu12} and the parameters $\mu_3,\mu_4$ are solutions of \eqref{eqn2} and given explicitly
in Eqs.~\eqref{mu01}-\eqref{mu34}. We will now determine the unknown coefficients $b_1,b_2,b_3,b_4$ and $b'_1,b'_2,b'_3,b'_4$ from the 
above conditions (i) and (ii) that we rewrite explicitly. The first one (i) is the continuity condition 
\be \label{condcont} 
b_1 V^1_{\sigma_1,\sigma_2} + b_2 
V^2_{\sigma_1,\sigma_2}(\mu_2) 
+ b_3 V^3_{\sigma_1,\sigma_2}(\mu_3) 
+ b_4 V^4_{\sigma_1,\sigma_2}(\mu_4) = 
b'_1 V^1_{\sigma_1,\sigma_2} + b'_2 
V^2_{\sigma_1,\sigma_2}(- \mu_2) +  
b'_3 V^3_{\sigma_1,\sigma_2}(- \mu_3) +
b'_4 V^4_{\sigma_1,\sigma_2}(- \mu_4) 
\ee 
and the second one (ii) is the matching of the derivatives, which reads
\bea \label{condder}
&& \mu_2 \left( b_1 V^1_{\sigma_1,\sigma_2} + b_2 
V^2_{\sigma_1,\sigma_2}(\mu_2) )  +
b'_1 V^1_{\sigma_1,\sigma_2} + b'_2 
V^2_{\sigma_1,\sigma_2}(- \mu_2) \right)
+ \mu_3 \left( b_3 V^3_{\sigma_1,\sigma_2}(\mu_3) + 
b'_3 V^3_{\sigma_1,\sigma_2}(- \mu_3) \right)
\\
&& + \mu_4 \left( b_4 V^4_{\sigma_1,\sigma_2}(\mu_4) + 
b'_4 V^4_{\sigma_1,\sigma_2}(- \mu_4) \right) = 
 2 \frac{\bar c}{D}  \times \left(
b_1 V^1_{\sigma_1,\sigma_2} + b_2 
V^2_{\sigma_1,\sigma_2}(\mu_2) 
+ b_3 V^3_{\sigma_1,\sigma_2}(\mu_3) 
+ b_4 V^4_{\sigma_1,\sigma_2}(\mu_4) \right) \;. \nn
\eea 
Finally we will also use that the total probability is normalized to unity. 
We note that $V^1_{\sigma_1,\sigma_2}$ is orthogonal to all the other vectors
for any value of $\mu$. We can thus take the scalar product of all equations with $V^1$ and obtain
\be \label{rel1}
b_1=b'_1 \quad , \quad \mu_2 (b_1+b'_1)=  2 \frac{\bar c}{D} b_1 \quad  \quad \quad \Longleftrightarrow \quad \quad b_1=b'_1=0 \;,
\ee 
as soon as $\gamma>0$, which is similar to our result for $p_2=P_{++}-P_{--}=0$ in the case $D=0$,
see \eqref{argument1}.

Based on some numerical observation, we now assume that
\be \label{rel2} 
b_2'=- b_2 \quad , \quad b_3'=b_3 \quad , \quad b_4'= b_4 \;,
\ee 
and we will verify below that it indeed provides a solution to the problem.
These identities imply from \eqref{condcont} that
\bea \label{condcont2} 
b_2 
(V^2_{\sigma_1,\sigma_2}(\mu_2) + V^2_{\sigma_1,\sigma_2}(- \mu_2)) 
+ b_3 (V^3_{\sigma_1,\sigma_2}(\mu_3) - V^3_{\sigma_1,\sigma_2}(-\mu_3))
+ b_4 (V^4_{\sigma_1,\sigma_2}(\mu_4) - V^4_{\sigma_1,\sigma_2}(-\mu_4))
= 0 \;,
\eea 
and from \eqref{condder} that
\bea \label{condder2} 
&& \mu_2 b_2 ( V^2_{\sigma_1,\sigma_2}(\mu_2) )  -  
V^2_{\sigma_1,\sigma_2}(- \mu_2) ) 
+ \mu_3 b_3 (V^3_{\sigma_1,\sigma_2}(\mu_3) + 
V^3_{\sigma_1,\sigma_2}(- \mu_3) )
\\
&& + \mu_4 b_4 ( V^4_{\sigma_1,\sigma_2}(\mu_4) + 
V^4_{\sigma_1,\sigma_2}(- \mu_4) ) = 
 2 \frac{\bar c}{D}  \times \left(b_2 
V^2_{\sigma_1,\sigma_2}(\mu_2) 
+ b_3 V^3_{\sigma_1,\sigma_2}(\mu_3) 
+ b_4 V^4_{\sigma_1,\sigma_2}(\mu_4) \right) \;. \nn
\eea 
Replacing the r.h.s. by half the sum of both sides of \eqref{condcont} and using
the above relations \eqref{rel1} and \eqref{rel2} one can rewrite \eqref{condder2} as
\bea \label{condnew}
&& (\mu_2 - \frac{\bar c}{D} ) b_2  ( V^2_{\sigma_1,\sigma_2}(\mu_2) )  -  
V^2_{\sigma_1,\sigma_2}(- \mu_2) ) 
+ (\mu_3 - \frac{\bar c}{D} )  b_3 (V^3_{\sigma_1,\sigma_2}(\mu_3) + 
V^3_{\sigma_1,\sigma_2}(- \mu_3) )
\\
&& + (\mu_4 - \frac{\bar c}{D} )  b_4 ( V^4_{\sigma_1,\sigma_2}(\mu_4) + 
V^4_{\sigma_1,\sigma_2}(- \mu_4) ) = 0 \nn
\eea 
Let us now express these equations using the explicit forms 
\bea \label{vectors} 
V^2(\mu_2) = \left(\begin{array}{c}
 \frac{c_2}{\sqrt{2}}  \\
 \frac{s_2}{\sqrt{2}}    \\
 -\frac{s_2}{\sqrt{2}}   \\
 \frac{c_2}{\sqrt{2}}   \\
\end{array} \right) \quad , \quad 
V^3(\mu_3) = \left(
\begin{array}{c}
  \frac{s_3}{2}  \\
  - \frac{1}{2} (1+ c_3)   \\
  - \frac{1}{2} (1- c_3)   \\
  \frac{s_3}{2}  \\
\end{array}
\right) 
\quad , \quad 
V^4(\mu_4) = \left(
\begin{array}{c}
  \frac{s_4}{2}  \\
   \frac{1}{2} (1- c_4)   \\
   \frac{1}{2} (1+ c_4)   \\
  \frac{s_4}{2}  \\
\end{array}
\right) \;,
\eea 
where, for $j=2,3,4$ one has
\be 
c_j = \frac{- \mu_j v_0}{\sqrt{\gamma^2 + \mu_j^2 v_0^2} }
\quad , \quad s_j = \frac{\gamma}{\sqrt{\gamma^2 + \mu_j^2 v_0^2}} \;,
\ee 
and $V^j(- \mu_j)$ obeys the same formula (\ref{vectors}) with $c_j \to - c_j$ and $s_j \to s_j$.
Inserting these expressions for $V^j(\pm \mu_j)$ into \eqref{condcont2} we obtain
only one independent equation 
\be  \label{eqnew1} 
b_3 c_3+b_4 c_4 = \sqrt{2} b_2 s_2 \;.
\ee 
Similarly, inserting them into \eqref{condnew} we obtain only two independent equations
\bea
&& b_4 \left(\frac{\bar c}{D}-\mu _4\right) = b_3 \left(\frac{\bar c}{D}-\mu _3\right)  \label{eqnew2}  \\
&& \sqrt{2} b_2 c_2 \left(\frac{\bar c}{D}-\mu _2\right)+b_3 s_3 \left(\frac{\bar c}{D}-\mu _3\right)+b_4 s_4 \left(\frac{\bar c}{D}-\mu _4\right) = 0 \;. \label{eqnew3} 
\eea 

The equations \eqref{eqnew1}, \eqref{eqnew2} give
\bea \label{b34} 
&& b_3 = \sqrt{2} b_2 \tilde b_3 \quad , \quad b_4 = \sqrt{2} b_2 \tilde b_4 \\
&& \tilde b_3= \frac{s_2 \left(\bar c- D \mu
   _4\right)}{\left(c_3+c_4\right) \bar c -D \left(c_4 \mu _3+c_3 \mu
   _4\right)} \quad , \quad \tilde b_4=  \frac{ s_2 \left(\bar c-D \mu
   _3\right)}{\left(c_3+c_4\right) \bar c - D \left(c_4 \mu _3+c_3 \mu
   _4\right)} \;,
\eea 
in which case we have checked that the third equation (\ref{eqnew3}) is automatically satisfied.

Putting everything together we find that the stationary measure is
\bea \label{statnew2} 
&& P_{\sigma_1,\sigma_2}(y) = b_2 \bigg(  (
V^2_{\sigma_1,\sigma_2}(\mu_2)  \theta(y) - 
V^2_{\sigma_1,\sigma_2}(- \mu_2) \theta(-y) ]
e^{-\mu_2 |y| }  \\
&& +  \tilde b_3 \sqrt{2} ( V^3_{\sigma_1,\sigma_2}(\mu_3) \theta(y) + 
V^3_{\sigma_1,\sigma_2}(- \mu_3) \theta(-y) ) e^{-\mu_3 |y|} 
+  \tilde b_4  \sqrt{2} (V^4_{\sigma_1,\sigma_2}(\mu_4) \theta(y) +
V^4_{\sigma_1,\sigma_2}(- \mu_4) \theta(-y) ) e^{-\mu_4 |y|} \bigg) \;,\nn
\eea
where $\tilde b_3,\tilde b_4$ are given in \eqref{b34} and there remains a single unknown parameter $b_2$ which is
obtained by normalization. Let us thus study the total probability $P(y)=\sum_{\sigma_1,\sigma_2} P_{\sigma_1,\sigma_2}(y)$. One has, from \eqref{vectors} 
\bea
 \sum_{\sigma_1,\sigma_2} V^2_{\sigma_1,\sigma_2}(\pm \mu_2)  = \pm \sqrt{2} c_2 \quad , \quad  \sum_{\sigma_1,\sigma_2} V^3_{\sigma_1,\sigma_2}(\pm \mu_3)  = s_3-1 
\quad , \quad \sum_{\sigma_1,\sigma_2} V^4_{\sigma_1,\sigma_2}(\pm \mu_4)  = s_4+1 \;.
\eea 
From \eqref{statnew2} we thus obtain 
\be \label{totP2} 
P(y) = b_2 \sqrt{2} \left( c_2  e^{- \mu_2 |y|} + \tilde b_3 (s_3-1)  e^{- \mu_3 |y|} + \tilde b_4 (s_4+1)  e^{- \mu_4 |y|} \right) \;,
\ee 
and the normalization condition $\int_{-\infty}^{+\infty} P(y) dy=1$ leads to the following result for $b_2$
\be 
b_2 = \frac{1}{\sqrt{2} \left( c_2  \frac{2}{\mu_2} + \tilde b_3 (s_3-1) \frac{2}{\mu_3}  + \tilde b_4 (s_4+1)  \frac{2}{\mu_4} \right)}  \;.
\ee 
Solving for $b_2$ and inserting into \eqref{totP2} we obtain the result for $P(y)$ given in the text in Eq. (\ref{fullP_Dpos}). Substituting into 
\eqref{statnew2} gives the complete result for all components of the stationary probability $P_{\sigma_1,\sigma_2}(y)$.
\\

{}


\begin{thebibliography}{}




\bibitem{soft} M. C. Marchetti, J. F. Joanny, S. Ramaswamy, T. B. Liverpool, J. Prost, M. Rao, and R. Aditi Simha, Rev. Mod. Phys. {\bf 85}, 1143  (2013).

\bibitem{BechingerRev} C. Bechinger, R. Di Leonardo, H. L\"{o}wen, C. Reichhardt, G. Volpe, and G. Volpe, Rev. Mod. Phys. {\bf 88}, 045006 (2016).

\bibitem{Ramaswamy2017} S. Ramaswamy, J. Stat. Mech.  054002, (2017).

\bibitem{Marchetti2018} \'{E}. Fodor, and M. C. Marchetti, Physica A {\bf 504}, 106 (2018). 

\bibitem{Berg2004} {\it E. Coli in Motion}, H. C. Berg,  (Springer Verlag, Heidel-
berg, Germany) (2004).

\bibitem{Cates2012} M. E. Cates, Rep. Prog. Phys. {\bf 75}, 042601 (2012). 

\bibitem{TailleurCates} 
J. Tailleur, and M. E. Cates, Phys. Rev. Lett. {\bf 100}, 218103 (2008).

\bibitem{kac74} M. Kac, Rocky Mountain J. Math. {\bf 4}, 497 (1974).

\bibitem{Orshinger90} E. Orsingher, Stoch. Process. Their Appl.
{\bf 34}, 49 (1990).

\bibitem{W02} G. H. Weiss, Physica A {\bf 311}, 381 (2002).

\bibitem{HJ95} P. H\"anggi and P. Jung, Adv. Chem. Phys. {\bf 89}, 239 (1995).



\bibitem{ML17} J. Masoliver and K. Lindenberg, Eur. Phys. J. B {\bf 90}, 1 (2017).




\bibitem{MJK18} K. Malakar, V. Jemseena, A. Kundu, K. V. Kumar, 
S. Sabhapandit, S. N. Majumdar, S. Redner, and 
A. Dhar, J. Stat. Mech. 043215, (2018).





\bibitem{Solon15} A. P. Solon, Y. Fily, A. Baskaran, 
M. E. Cates, Y. Kafri, M. Kardar, and J. Tailleur, 
Nature Phys. {\bf 11}, 673 (2015).

\bibitem{TDV16} S. C. Takatori, R. De Dier, J. Vermant, and J. F. Brady, Nature Comm. {\bf 7}, 10694 (2016).

\bibitem{DKM19} 
A. Dhar, A. Kundu, S. N. Majumdar, S. Sabhapandit and 
G. Schehr, Phys. Rev. E {\bf 99}, 032132 (2019).


\bibitem{BMR19} U. Basu, S. N. Majumdar, A. Rosso, and G. Schehr, Phys. Rev. E, {\bf 100}, 062116 (2019). 

\bibitem{DD19} O. Dauchot and V. D\'emery, Phys. Rev. Lett. {\bf 122}, 068002 (2019).



\bibitem{3statesBasu}
U. Basu, S. N. Majumdar, A. Rosso, S. Sabhapandit, and G. Schehr, J. Phys. A: Math. Theor. {\bf 53}, 09LT01 (2020)

\bibitem{LMS2020}
P. Le Doussal, S. N. Majumdar, and G Schehr, EPL {\bf 130}, 40002 (2020).





\bibitem{slowman}
A. B. Slowman, M. R. Evans, and R. A. Blythe,  Phys. Rev. Lett. {\bf 116}, 218101 (2016).

\bibitem{slowman2}
A. B. Slowman, M. R. Evans, and R. A. Blythe,  J. Phys. A: Math. Theor. {\bf 50}, 375601 (2017).


\bibitem{MBE2019}
E. Mallmin, R. A. Blythe, and M. R. Evans, J. Stat. Mech., 013204 (2019).



\bibitem{KunduGap2020}
A. Das, A. Dhar, and A. Kundu, J. Phys. A: Math. Theor. {\bf 53}, 345003 (2020).

\bibitem{LMS2019}
P. Le Doussal, S. N. Majumdar, and G. Schehr,  Phys. Rev. E {\bf 100}, 012113 (2019).













\bibitem{cardano_wiki}
\url{https://en.wikipedia.org/wiki/Cubic_equation}.

\bibitem{SBS2020}
I. Santra, U. Basu, S. Sabhapandit, Phys. Rev. E {\bf 101}, 062120 (2020).




\bibitem{SinghChain2020} 
P. Singh, A. Kundu, preprint arXiv:2012.13910.


\bibitem{PutBerxVanderzande2019}
S. Put, J. Berx, and Carlo Vanderzande, J. Stat. Mech. 123205 (2019).


\bibitem{DMS21} 
D. S. Dean, S. N. Majumdar, and H. Schawe,  Phys. Rev. E {\bf 103}, 012130 (2021).

\end{thebibliography}
\end{document}